# Atomistic insights into hydrogen bonds of water around detonation nanodiamonds: effects of surface chemistry and dissolved ions

Farshad Saberi-Movahed* and Donald W Brenner

Department of Materials Science and Engineering, North Carolina State University, Raleigh, NC, USA.

(*Corresponding author: fsaberi@ncsu.edu)

**ABSTRACT**

Prior studies on the effects of surface chemistry of detonation nanodiamonds (DNDs) on Hydrogen Bonds (HBs) of water have not unambiguously distinguished among three broad categories of HBs: water-site HBs ($HB^{(s)}$), water-water HBs in hydration layers of DNDs ($HB^{(h)}$), and water-water HBs in the bulk region ($HB^{(b)}$). Furthermore, the effects of dissolved ions in the solution on the aforementioned HBs have not been studied yet. In this study, we addressed these issues via Molecular Dynamics simulations. Each system under study contained a single DND that was functionalized either with purely –H or with a mixture of –H and one of –$NH_2$, –COOH, or –OH groups. Then, the product was dissolved in 0.1M aqueous solution of any of KCl, NaCl, $CaCl_2$, or $MgCl_2$ salts. All DNDs except for DND–OH, which was charge neutral, existed as either a charged or a neutral particle in the solution. Whereas DND–H with +56 net charges induced the strongest $HB^{(h)}$, the addition of +28 charges to DND–H led to a drop in $HB^{(h)}$ strength and became identical to what we observed for $HB^{(h)}$ strength around DND–OH. Considering only neutral DNDs, $HB^{(h)}$ strength increased in the sequence as DND–H ≈ DND–COOH < DND–OH < DND–$NH_2$. For the case of the negatively charged DND–COOH, while $HB^{(h)}$ in either NaCl or KCl solution was slightly weaker than $HB^{(b)}$, $HB^{(h)}$ strength corresponding to $CaCl_2$ or $MgCl_2$ solutions were enhanced compared to $HB^{(b)}$. Interestingly, $Mg^{2+}$ and to a lesser extent $Ca^{2+}$ around the charged DND–COOH were associated with the strongest $HB^{(s)}$ in this study.

**Keywords:** Nanodiamond; Functional Group; Ion; Hydration Shell; Hydrogen Bond; Structure-breaker (Chaotrope); Structure-maker (Kosmotrope); Specific Ion Effects; Molecular Dynamics.

## 1 Introduction

The interactions of water molecules with solid surfaces are ubiquitous in many environments such as the aqueous solution of proteins[1–4], DNAs[5–8], and nanoparticles (NP)[9–13]. These interactions, which can be modulated by some co-solvents and solvated ions[14], are shown[15] to have mutual effects on both solute and water. The mutual effects involve the functioning of the solute such as conformational changes and the solubility of biomolecules[16–18], on the one hand, and the physical properties of the aqueous solution such as the viscosity[14,19], diffusion[20], heat capacity[21], and freezing/melting point[22,23] on the other hand.

Numerous studies have shown that the aforementioned coupling effects manifest through modifications in hydrogen bond (HB) network of water induced by solute-water interactions.[24–30] Detonation Nanodiamond (DND) particle has been the subject of such studies,[15,31–36] due to its promising performance in some biomedical applications,[37–44] where DND-water interactions are inevitable.[34] Furthermore, some post-synthesis purification techniques–to improve the colloidal stability of DNDs and also to tune their surface chemistry for specific applications–take place in the aqueous environments.[45–47] The colloidal stability of DNDs and their fluorescent properties in the aqueous solutions are among the most investigated characteristics of DNDs that have been unequivocally related to strengths of HBs in water around DNDs.[15,31,34,48–54]

However, to the best of our knowledge, the effects of ions on HBs of water around DNDs have not been investigated, although the adsorption of different ions on carboxylated DNDs have been studied[55–58] before. Indeed, there is a wealth of research on the effects of dissolved ions on the HB network of water with concomitant consequences on physical and thermodynamical properties of aqueous solutions such as viscosity, solubility, and enthalpy of solution.[59–67] Depending on their weakening or strengthening effects on the HB network of water, ions are conventionally categorized into "structure-breaker" (chaotrope) or "structure-maker" (kosmotrope) species, respectively.[68,69] The first group of ions are characterized as relatively low charge density (usually large, monovalent) ions such as $Cs^+$, $K^+$ or $Cl^-$, whereas the second group comprises ions of opposite characteristics, that is, small, multivalent ions such as $Mg^{2+}$, $Ca^{2+}$, or polyatomic ions such as $CO_3^{2-}$ or $SO_4^{2-}$.[70] Thus, "structure-maker" ions can attract water molecules closer to themselves and exert stronger electrostatic forces on them than "structure-breaker" ions do.[71] Consequently, as opposed to the latter, remarkably higher structured water with extremely retarded dynamic features is organized around the former with respect to bulk water's organization.[72–74] However, some ions such as $Na^+$ are considered as a borderline case.[68] Because, although they can impart a moderately ordered structure in their first hydration shell, they fail to extend it to their second hydration shell and beyond.[75,76] Nevertheless, there is still a contentious debate over whether effects of ions on HBs of water extend from their hydration shells to bulk water.[75,77–83]

The effects of various surface functional groups of DNDs on the strength of water HBs around them are also not clear. Two earlier studies[15,31] proposed DND–H > DND–COOH > DND–polyFunc > DND–OH ordering in terms of the weakening effects of the surface functional groups (–H, –COOH, –polyFunc, –OH) on the strength of water HBs around DNDs. In contrast, a recent study[84] conducted by Laptinskiy et al. offered a completely different ordering as DND–H > DND–polyFunc > DND–OH > DND–COOH. Here, –polyFunc refers to a mixture of surface functional groups such as –OH, –COOH, –H, and so on. In all of these three studies as well as Ref.[15,31,32,48–54], the strength of water HBs have been inferred by identifying changes in the shape and position of features in the OH stretch band of the Raman spectroscopy of the DND aqueous suspensions. The low-frequency and high-frequency features of this band correspond to environments in the solution with, respectively, strong and weak HBs. Stehlick el al. used the OH stretch band in the Infrared (IR) spectroscopy instead of Raman spectroscopy as their main characterization tool.[54] They predicted that the strength of HBs between water and oxidized DNDs is weaker than that of bulk water-water HBs, although the exact type of oxygen containing functional groups in their study is unknown.

All abovementioned studies except Ref.[84] presented their conclusions for HBs of water based on qualitative trends in features of the OH stretch band against some other parameters such as concentration of DNDs or temperature. In contrast, Ref.[84] supported its claims based on the enthalpy change of HBs that were calculated from the van't Hoff equation. To carry out these calculations, they decomposed the OH stretch band into Gaussian-shaped components in order to quantify contributions of water OH groups with weak and strong HBs to the OH stretch band. Nevertheless, all of these studies suffer from a common shortcoming. None of them have made any clear distinctions among three main categories of HBs of water (see Figure 1), namely, water-water HBs in the bulk region ($HB^{(b)}$), water-water HBs in the hydration shells of the DND ($HB^{(h)}$), and HBs of water with surface functional groups of DNDs ($HB^{(s)}$). On the contrary, what they have presented is an average over contributions of these three types of water to the vibrational modes of water in Raman/IR spectroscopy. In fact, prior studies have provided some predictions for the strength of water-site HBs, which are mostly based on some heuristics such as the electronegativity of functional groups[15,48,54] or the amount of water adsorbed onto DND's surface[54]. The quantification of the strength of water-site HBs becomes particularly imperative in the light of the existing theory for the relationship between the DND's fluorescent intensity and the strength of DND-water HBs.

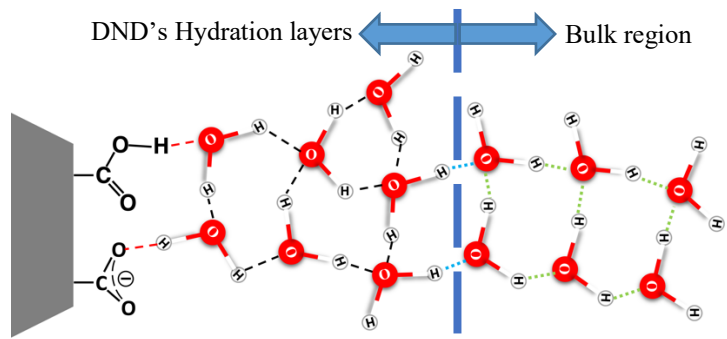

**Figure 1.** Illustration of three types of HBs of water in three different regions around DND. Red dashed, black dashed, and green dotted lines represent water-site HB ($HB^{(s)}$), water-water HB in the hydration layers of DND ($HB^{(h)}$), and water-water HB in the bulk region ($HB^{(b)}$).

The assignment of features in the OH stretch band recorded by Raman/RI spectroscopy to water molecules in different environments (in our case, bulk, hydration shell, and interface with surface functional groups of DND) with distinct HB strengths is not straightforward. First of all, this band contains vibrational responses from intermolecular coupling, intramolecular coupling (known as Fermi resonance), and HB interactions of liquid water.[85,86] To turn off the inter-/intra-coupling, the isotopic dilution of water is usually used.[87,88] Second, the OH stretch band of the aqueous solution of any solute needs to be processed to unveil contributions from water molecules with different strengths of HBs in hydration shells of solutes and in the bulk region. For this purpose, solute-correlated and bulk-correlated spectra are obtained by using such methods as multivariate curve resolution or peak fitting (with a presumed fitting functional form such as Gaussian).[84,89–91] However, each of these methods is subject to some kind of assumptions.[87] Another way to blur the effect of bulk water on the OH stretch band in order to obtain the spectra features of the solute's hydrating water is to disperse high concentrations of the solutes in water.[85] But this method is problematic for the aqueous solution of DNDs, as they are prone to aggregate.[15,48] Finally, it is possible that a certain vibrational frequency of the solute (here, DND) interferes with that of water in the Raman/IR spectra, which needs to be taken care of.[53]

In this paper, we set forth a journey to obtain a molecular-level understanding about how following factors affect HBs of water around a single cuboctahedral DND particle: 1) the type of functional groups on facets of the DND, 2) the amount of DND's net surface charges, and 3) the type of dissolved ions with different charge densities. To this end, we have employed all-atom Molecular Dynamics[92,93] (MD) simulations with explicit water, as an alternative approach to Raman/IR spectroscopic techniques.

MD simulations afford us not only the distinctions among $HB^{(s)}$, $HB^{(h)}$, and $HB^{(b)}$, but also a molecular picture of the underlying mechanism through which DNDs modify their nearby water's HBs. In the pursuit of this mechanism, we characterize the dynamics of different categories of HBs using time-dependent autocorrelation functions of two state variables.[94] The integrations of these two functions yield the lifetime of HBs before breaking (hence, a measure of HB's strength) and the structural relaxation time of the HB network.[95,96] The latter is related to the collective rearrangements of the HB network, resulting from coupled reorientational-translational motions of its constituent species.[96–100]

The rest of this paper is organized as follows. Section 2 provides the details of the model system for MD simulations, along with the computational methods for characterizing HBs. Then, Section S.3 presents some statistics about number of HBs per water molecules. Dynamic characteristics of water-water, water-site, and water-anion HBs constitute the subject of Section S.4. Finally, our concluding remarks are presented in Section 4.

## 2 Methodology

### S.1 MD simulation setup

In our MD simulations, a single cuboctahedral DND with *ca.* 4.4 nm in diameter is solvated in a cubic box of 0.1M aqueous solution of any of KCl, NaCl, $CaCl_2$, and $MgCl_2$ inorganic salts. We have considered four different types of surface functionalizations, namely fully hydrogenated (DND–H), hydroxylated (DND–OH), carboxylated (DND–COOH), and aminated (DND–$NH_2$). The last three types of DNDs also have hydrogen atoms in addition to the polar groups on their facets. While DND–OH is charge neutral in our study, other three types of DNDs exist with four different net charges with absolute values of 0, 28, 56, or 84. The charged versions of DND–H and DND–$NH_2$ have positive charges, whereas the charged DND–COOH has negative ones. Negative charges of DND–COOH and positive charges of DND–$NH_2$ result from the deprotonation of –COOH and the protonation of –$NH_2$ groups, respectively. Thus, the charged variants of DND–COOH and DND–$NH_2$ have, respectively, additional –$COO^-$ and –$NH_3^+$ groups, with respect to their corresponding uncharged counterparts. The net positive charges of DND–H are obtained from the modification of the partial charges (hence, the electronegativity) of surface C and H atoms through a computational procedure developed by Su *et al.*[101]. Overall, there are 52 distinct atomistic systems for each of which we ran five independent MD simulations to obtain a statistically reliable information from the simulations. We have described a step-by-step procedure to carry out these simulations in our previous work[58].

### S.2 Computational Tools

In order to post-process the atomic trajectories for characterizing HBs, we developed a series of computational tools as adds-on to MDAnalysis[102,103] package. Our python codes incorporate the Message Passing Interface (MPI) for parallel computing to accommodate heavy computations on massive trajectory data through distributed high-performance computing systems.

### 2.1.1 Hydrogen Bond Structure

There are different techniques reported in the literature to identify an HB between two molecules. We can classify them into two groups of energy-based and geometry-based techniques, based on different criteria of detecting HBs. While the former uses energetic criteria to detect HBs, the latter uses geometric criteria based on the relative spatial and orientational configurations of two neighbor molecules[104,105]. Both techniques can reproduce experimental predictions of number of HBs in water and can also yield almost identical characterization of HBs' dynamics[105]. However, we have employed the geometry-based technique, which is also widely used in the research community[106–109], due to its relative simplicity.

We consider two molecules are hydrogen bonded, as shown schematically in Figure 2(a), if the following geometric criteria are met[110]:

$$d_{D\cdots A} < d^c \qquad \text{Eq. 1}$$

$$\alpha_{A\cdots D-H} < \alpha^c \qquad \text{Eq. 2}$$

where D and H are, respectively, donor and hydrogen atoms in the molecule that donates the HB, A is the HB acceptor in the other molecule, $\alpha_{A\cdots D-H}$ is the HB angle, $d_{D\cdots A}$ is the donor-acceptor distance, $d^c$ is the cut-off distance, and $\alpha^c$ is the cut-off angle. The value for $d^c$ is usually determined from the first minimum in the partial RDF of $D-A$. For instance, the distance and angle cut-off values for water-water HB are 3.5 Å and 30°, respectively.[106,107,111]

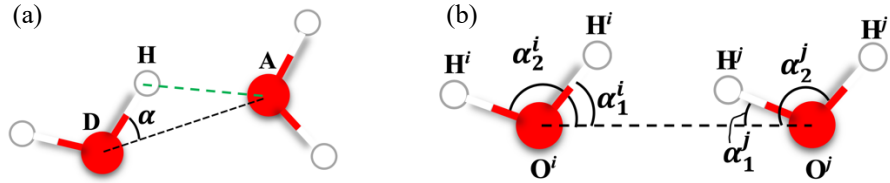

**Figure 2.** (a) Geometry-based definition of HB, where two water molecules are hydrogen bonded if the distance between donor D and acceptor A atoms ($d_{D\cdots A}$) and $\alpha$ angle satisfy geometric criteria in Eq. 1 and Eq. 2 respectively, (b) demonstration of four angles that are formed between one of OH bonds of either water molecules and the connector between their oxygen atoms. Red and white particles are oxygen and hydrogen atoms of water.

In practice, the identification of the donor and acceptor atoms can be slightly complicated. Let's assume that the distance between $O^i$ and $O^j$ atoms of two water molecules, as shown in Figure 2(b), meets the criterion in Eq. 1. Then, we are faced with four angles, i.e., $\alpha_1^i$, $\alpha_2^i$, $\alpha_1^j$, and $\alpha_2^j$, of which one, two, or even more might meet the criterion in Eq. 2. However, only one of those angles could be the right HB angle. As suggested by Raschke *et. al.*[112], we choose the smallest of those angles as the HB angle that also meets the criterion in Eq. 2. Consequently, if there is such an angle, then the associated water molecule is considered the HB donor and the other water molecule is tagged as the HB acceptor. This approach can be applied to any pair of molecules that potentially can form an HB, although the number of candidate angles for the right HB angle could be different from what we discussed above.

### 2.1.2 Hydrogen Bond Dynamics

Many experimental and computational studies, for instance Ref.[113], have reported that HB network of water is not static. In contrast, it is labile and undergoes continuous rearrangements. Therefore, it is imperative to characterize the dynamics of these rearrangements for water molecules in the vicinity of solutes, given the importance of HBs in the solute-water interactions[96]. To this end, we have employed two widely used HB time correlation functions, of which Rapaport[94] was one of the first pioneers. These are the intermittent $C_{HB}^{(i)}(t)$ and the continuous $C_{HB}^{(c)}(t)$ time correlation functions, referred to as ITCF and CTCF respectively, and may be written as[114]:

$$C_{HB}^{(i)}(t) = \frac{\langle h(t_0).h(t)\rangle}{\langle h(t_0)\rangle} \qquad \text{Eq. 3}$$

$$C_{HB}^{(c)}(t) = \frac{\langle h(t_0).H(t)\rangle}{\langle h(t_0)\rangle} \qquad \text{Eq. 4}$$

where $h(t)$ and $H(t)$ are both HB population operators with values of 0 or 1, although their evaluations are different. $H(t)$ is 1 if a tagged pair of water molecules continuously remains hydrogen bonded from $t_0$ to $t$, otherwise it is evaluated to 0. In contrast, $h(t)$ is unity as long as two molecules are hydrogen bonded at both $t_0$ and $t$ times, regardless of whether the HB was broken and reformed again in the interim. We should emphasize that $\langle \cdots \rangle$ denotes an ensemble average over only pairs of molecules that are hydrogen bonded at $t_0$. Here, $t_0$ represents different time origins. While $C_{HB}^{(c)}(t)$ gives an estimate of the HB strength, its intermittent counterpart provides a measure of the survival probability of an initial HB between two molecules.[105] Furthermore, the time integration of $C_{HB}^{(c)}(t)$ provides the average lifetime of HB ($\tau^{(c)}$), whereas that of $C_{HB}^{(i)}(t)$ reflects the structural relaxation of the HB network, denoted as $\tau^{(i)}$.[105]

## 3    Results and Discussions

### S.3    Statistics of HBs' population

Figure 3 shows the average number of water-water HBs ($\overline{N}_{HB}$) per each water molecule in the whole hydration shell of various DNDs that are solvated in different salt solutions. The numerical values of all statistics related to the number of HBs are listed in the Supplementary Information (SI) (Table S.1 to Table S.6). The whole hydration shell of a DND is defined as a 1.0 nm thick spherical shell with an inner radius of ca. 1.8 nm that is centered at the centroid of the DND. We have determined the whole hydration shell's thickness from the Perpendicular Number Density (PND) plots provided elsewhere[58]. For comparison, Figure 3 also includes $\overline{N}_{HB}$ values for bulk water. Throughout this study, bulk water refers to water molecules in the outside region of the whole hydration shell of the neutral DND–H in each of salt solutions– that is, far away from the influence of DNDs' surfaces.

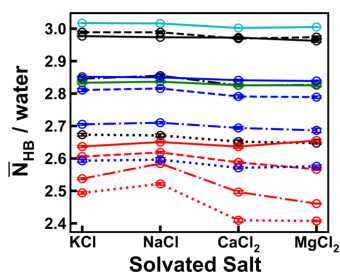

**Figure 3.** The average number of water-water HBs per each water molecule ($\overline{N}_{HB}$ / water) in the whole hydration shell of different DNDs with various surface charges (differentiated by different line styles) in different salt solutions. Black, blue, red, and green lines represent DND–H, DND–NH$_2$, DND–COOH, and DND–OH particles, respectively, and the cyan line shows the values of bulk water. Solid, dashed, dash-dotted, and dotted lines correspond to 0, 28, 56, and 84 absolute charges on DNDs, respectively. DND–H and DND–NH$_2$ assume any of these charges with the positive sign, so does DND–COOH but with the negative sign. However, DND–OH only exists as a neutral particle in our study. Disclaimer: the lines are just included for visual aid, as there is no continuity between categories on the x-axis!

We can see in Figure 3 that $\overline{N}_{HB}$ values of bulk water in all salt solutions are almost identical and equal to 3, which is slightly smaller than the reported value of 3.3 for pure water.[68] We attribute the difference to the effect of solvated ions, which disrupt the HB network of water.[68] Compared with all other DNDs, corresponding $\overline{N}_{HB}$ values to DND–H with 0 and +28 charges in all salt solutions are the closest to those of bulk water. This can be explained by the hydrophobic nature of DND–H, where the interfacial water does not form any HBs with facets of the DND.

However, as DND–H acquires more positive charges, $\overline{N}_{HB}$ decreases. In the extreme case of DND–H with +84 charges, $\overline{N}_{HB}$ drops below that of the neutral DND–H by 11%. The reduction in $\overline{N}_{HB}$ can be explained by positive charges of DND–H that have a twofold bearing. First, the positively charged DND–H attracts water oxygen closer to and repel water hydrogens away from its facets.[115–118] It results in the loss of two HBs that the water oxygen could have accepted from neighboring water molecules. Second, Cl⁻ anions are adsorbed onto facets of the positively charged DND–H, each of which can accept an HB from the interfacial water. As a reminder, a Cl⁻ anion is hydrated by at most seven water molecules with one of their OH bonds oriented toward the anion.[68,73] Thus, an adsorbed Cl⁻ can cause at most seven water molecules at the interface with the positively charged DND–H to each lose one HB.

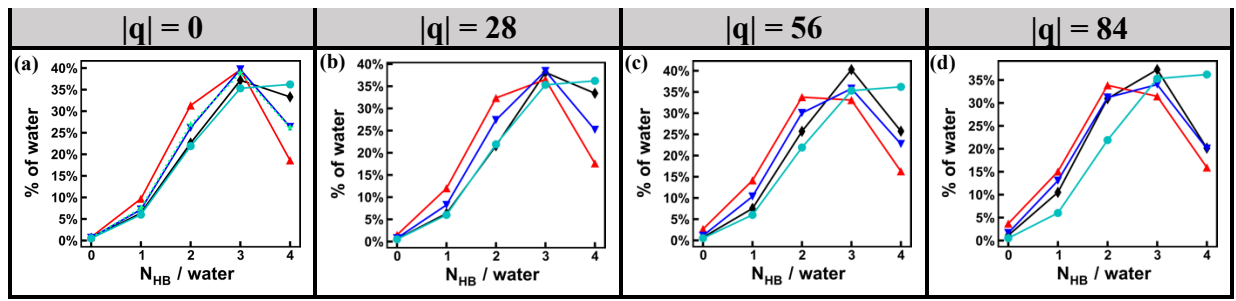

**Figure 4.** The distribution of number of water-water HBs per each water molecule in the whole hydration shell of different DNDs with various surface chemistries in MgCl$_2$ aqueous solution, where |q| represents the absolute charges on DNDs as described in Figure 3. Black, blue, and red solid lines and the green dotted line (part (a)) represent DND–H, DND–NH$_2$, DND–COOH, and DND–OH particles, respectively, while the cyan solid line corresponds to bulk water.

Both aforementioned effects, which are also reflected in Figure 4, are more pronounced in the case of DND–H with +84 charges.[68] Indeed, Figure 4 reveals that 33%, 33%, 25%, and 20% of water molecules in the whole hydration shell of DND–H with, respectively, 0, +28, +56, and +84 charges have 4 water-water HBs. Thus, we observe that the relative population of tetrahedrally coordinated water molecules drops by 13%, as the neutral DND–H obtains +84 net charges. We ascribe this effect to adsorbed Cl⁻ anions onto facets of the latter, which in turn disrupt water's HB network as we described above. On the other hand, DND–H with +84 charges is associated with 11% and 31% of water molecules with, respectively, 1 and 2 water-water HBs, indicating 5% and 8% jumps with respect to the corresponding values of the neutral DND–H. The 5% boost is indicative of the combined effect of the adsorbed Cl⁻ anions and the preferential orientation of water's OH bonds around DND–H with +84 charges. Furthermore, Figure 4 shows that 40% of water molecules in the whole hydration shell of DND–H with +56 charges have 3 water-water HBs, which is the highest percentage compared with other DND–H particles. It points to an enhancement in water-water HB formation in the first hydration layer of DND–H with +56 charges; it results from those water molecules having one of their OH bonds almost parallel to a facet in order to donate a HB to a neighboring water.

Interestingly, Figure 3 reveals almost equal values of $\overline{N}_{HB}$ for both DND–OH and the neutral DND–NH$_2$. We also observe in Figure 4(a) almost identical percentages of water with a specific number of water-water HBs in the whole hydration shell of these DNDs. It suggests that there exists a similar hydrogen bonding mechanism in water nearby DND–OH and the neutral DND–NH$_2$. Based on our previous observations,[69] we note that both –OH group on the former and –NH$_2$ group on the latter induce almost identical preferential orientations in the respective DND's interfacial water. More specifically, there are two groups of water, W1 and W2, at the interface with these DNDs that exhibit different orientations as follows: 1) W1 water orients one of its OH bonds toward its nearby facet to donate an HB to –OH or –NH$_2$ groups, 2) W2 water points its oxygen toward and its OH bonds away from the facet to accept a HB from aforementioned groups. The second orientation is more prevalent on {111} facets than it is on {100} facets. Thus, W1 and W2 water lose one and two water-water HBs, respectively, which is also reflected in Figure 4(a). That is, we observe 4% increase in the percentage of water molecules with either 2 or 3 water-water HBs nearby DND–OH and the neutral DND–NH$_2$, with respect to those in bulk water. In contrast, the percentage of water molecules with 4 HBs declines from 36% in bulk water to 26% in the whole hydration shell of either of these DNDs.

Positively charged DND–NH$_2$, whose facets are decorated with $-NH_3^+$ group in addition to –NH$_2$ and –H groups, exhibits some disruptions in the abovementioned HB patterns of its neutral counterpart. In particular, water surrounding DND–NH$_2$ with +84 charges has, on average, 7% less water-water HBs than that nearby the neutral DND–NH$_2$ (see Figure 3). Figure 4(d) gives a clue about the origins of the aforementioned difference. The percentages of water molecules with 1 and 2 water-water HBs in the whole hydration shell of DND–NH$_2$ with +84 charges have increased by, respectively, 6% and 5% with respect to those of the neutral DND–NH$_2$. We ascribe the increase in water molecules with 2 water-water HBs to the preferential orientations of water, induced by $-NH_3^+$ groups, in the immediate vicinity of DND–NH$_2$ with +84 charges. $-NH_3^+$ groups push OH bonds of the interfacial water away from and instead attract its oxygen toward a nearby facet in order to donate a HB to it.[68] Thus, it leads to a loss of 2 water-water HBs for water molecules in the vicinity of DND–NH$_2$ with +84 charges. In addition, adsorbed Cl$^-$ anions can cause some of these water molecules to lose a third water-water HB through a similar mechanism we described previously for DND–H with +84 charges. Thus, it explains the aforementioned 6% boost in the parentage of water molecules with 1 water-water HB around DND–NH$_2$ with +84 charges.

Last but not least, we address water-water HB statistics in the whole hydration shell of DND–COOH. In this regard, we identify two interesting features in Figure 3. First, $\overline{N}_{HB}$ values of DND–COOH are smaller than those of other DNDs, for each specific net absolute charge (|q|) of the DND and the type of the salt solution. For instance, water in the whole hydration shell of the neutral DND–COOH has, on average, 11%, 7%, and 7%, less water-water HBs than that of the neutral DND–H, the neutral DND–NH$_2$, and DND–OH, respectively. Moreover, water in the whole hydration shell of DND–COOH with –84 charges solvated in either of CaCl$_2$ or MgCl$_2$ salt solutions has, on average, 2.4 HBs with neighboring water molecules. It is the smallest number in our study, which is 20% less than $\overline{N}_{HB}$ in bulk water. Second, we observe the specific cation effect on $\overline{N}_{HB}$ values of charged DND–COOH, particularly for those of the DND with –56 and –84 charges. That is, $\overline{N}_{HB}$ variations depend on the type of the solvated cation in the solution. We expand on each of these features below.

Figure 4, which is for MgCl$_2$ solution, helps us dig deeper to find out why $\overline{N}_{HB}$ values of DND–COOH are lower than those of other DNDs with the same number of absolute charges. It shows for various values of |q| that percentages of water molecules with 1 and 2 water-water HBs in the surrounding of DND–COOH are larger than those of other DNDs. The reverse trend is true for percentages of water molecules having 3 (except for |q| = 0) and 4 water-water HBs. In the extreme case of |q| = 84, there exist water molecules, making up around 4% of the whole hydration shell's population, that do not have any HBs with other water molecules in the surrounding of DND–COOH. This number amounts to roughly 1% and 2% for, respectively, DND–H and DND–NH$_2$ with +84 charges on each.

We ascribe abovementioned trends for DND–COOH to disruptions in water-water HBs that are made by –COOH and –COO⁻ on the one hand and adsorbed cations (for charged DND–COOH) on the other hand. Water in the immediate vicinity of the neutral DND–COOH, that has only –COOH and –H groups on its facets, adopts two different orientations.[58] Water near the hydroxyl part of –COOH group orients its dipole away from the nearby facet in order to accept a HB from the hydroxyl. It leads to a loss of at least two water-water HBs for each of these water molecules. On the other hand, the doubly bonded oxygen in –COOH (hereafter, O2) attracts the nearby water's hydrogen, which forces water to orient one of its OH bonds and thereby its dipole moment toward the associated facet. Although O2 atom does not form a strong HB with the interfacial water (as we will see later), it causes its approaching water to lose at least 1 water-water HB. However, the deprotonation of sufficiently large number (ca. 56 groups or more) of –COOH groups on facets of the neutral DND–COOH leads to the domination of the second type of aforementioned orientations over the first one.[58] In other words, the resulting –COO⁻ groups on the negatively charged DND–COOH are greedy to accept HBs from the interfacial water. Thus, they entice water molecules to orient their OH bonds toward their nearby facets to donate HBs to oxygens of –COO⁻ groups.

Cations that are adsorbed near –COO⁻ groups on the charged DND–COOH can have a compound effect on hydrogen bonding behavior of the interfacial water. We found in our previous work[58] that different cations exhibit distinct adsorption behavior, which depends on their charge density and hence their water affinity. Among four cations, Na⁺ has the highest affinity for –COO⁻ anion and predominantly forms the Contact Ion-Pair (CIP) association with it. The majority of adsorbed K⁺ cations also form CIP associations with –COO⁻ group, although their interaction strengths are weaker than those of Na⁺–COO⁻. In contrast, the Solvent-shared Ion-Pair (SIP) is the dominant complexation between $Mg^{2+}$ and –COO⁻ anion. Adsorbed $Ca^{2+}$ cations predominantly form CIP complexations with –COO⁻ anion, although their SIP associations become gradually noticeable on DND–COOH with –56 and –84 charges. In the CIP association, a cation and an anion shed some of their hydrating water and come to a close contact. In contrast, the SIP association arises from oppositely charged particles sharing a water molecule in their respective first hydration shells.

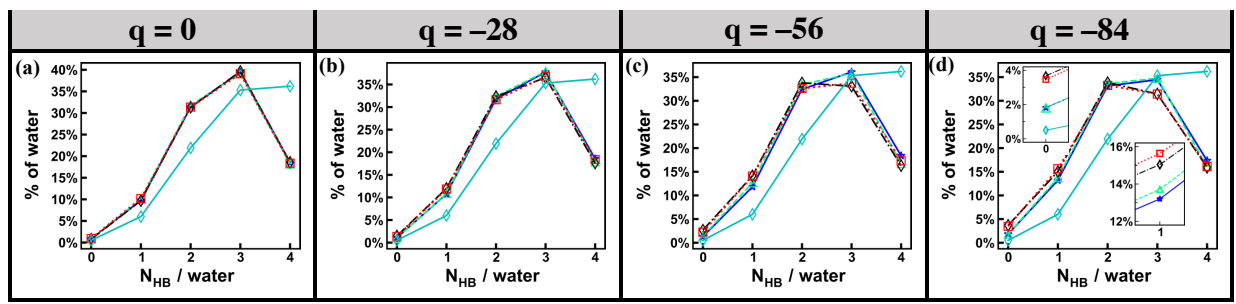

**Figure 5.** The distribution of number of water-water HBs per each water molecule in the whole hydration shell of DND–COOH with various charges (q), solvated in each of KCl (dashed green line), NaCl (solid blue line), $CaCl_2$ (dotted red line), and $MgCl_2$ (dash-dotted black line) aqueous solutions. Solid cyan line represents similar values but for bulk water.

As we can see in Figure 5(d), COOH–DND with –84 charges in solutions containing divalent cations are associated with slightly higher percentages of water molecules with 0 and 1 water-water HBs than solutions with monovalent cations. Conversely, we observe a reverse trend for the percentage of water molecules with 3 water-water HBs. Figure 5(d) also reveals similar trends for COOH–DND with –56 charges, yet with lesser intensities. We attribute the first trend to the CIP complexation of each of $Mg^{2+}$ and $Ca^{2+}$ cations with $-COO^-$ surface-bound anion. The second trend arises from more strongly bound first hydration shell of $Mg^{2+}$ and $Ca^{2+}$ cations than that of $Na^+$ and $K^+$ cations. Thus, water in the first hydration shell of the monovalent cations have more rotational freedom to form HBs with water in their second hydration shell.[58,68,70,73,119] We will expand more on this topic later in the present paper.

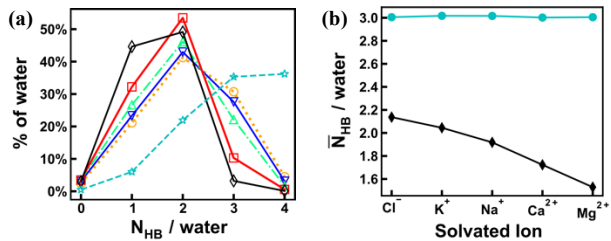

**Figure 6.** Statistics of water-water HBs in the whole hydration shell (i.e., the combined first and second shells) of solvated ions in four different aqueous salt solutions of the neutral DND–H; (a) the distribution of number of water-water HBs per each water molecule, (b) the average number of water-water HBs per each water molecule. In part (a), solid blue, red, and black lines and dash-dotted green, dotted orange, and dash cyan lines correspond to water around $Na^+$, $Ca^{2+}$, $Mg^{2+}$, $K^+$, $Cl^-$ (of $MgCl_2$ salt) ions and water in bulk, respectively. In part (b), black line with diamond markers and cyan line with circle markers represent, respectively, values around ions and values corresponding to the bulk water.

We have also calculated $N_{HB}$/water distributions and $\bar{N}_{HB}$ values for water in the whole hydration shell (i.e., the combined first and second shells) of solvated ions, which are shown in Figure 6. The results in this figure are obtained from the aqueous solution of the neutral DND–H data, where ions are less likely to adsorb onto facets of the DND. Strikingly, water around $Mg^{2+}$ has the least average number of HBs with its neighboring water molecules, compared with that around other ions (see Figure 6(b)). It is the signature of the remarkably high charge density of $Mg^{2+}$, which leads to pronounced restrictions on reorientations of its first hydration shell water's dipole and OH bonds.[58,119] Thus, water in the first hydration shell of $Mg^{2+}$ forms fewer HBs with water in its second hydration shell, which is also evident from the distribution of HBs in Figure 6 (a). A similar effect, yet to a lesser degree, takes place in water around $Ca^{2+}$. On the other hand, owing to smaller charge densities of other three ions, their surrounding water is much less restricted and thus can form more HBs with nearby water molecules. Furthermore, smaller $\bar{N}_{HB}$ values corresponding to $Ca^{2+}$ and $Mg^{2+}$ relative to $K^+$ and $Na^+$ can explain the smaller $\bar{N}_{HB}$ values for water-water HBs in the whole hydration shell of DND–COOH with –84 charges in $CaCl_2$ and $MgCl_2$ solution (see Figure 3).

**S.4  Dynamics of HBs**

In this section, we investigate the dynamics of HBs that water molecules form with other water molecules (water–water), $Cl^-$ anion (water–anion), and polar functional groups on DNDs (water–site HB). The first one corresponds to water molecules that reside in hydration shells of DNDs, cations, and $Cl^-$ anion. To this end, we have employed CTCF and ITCF functions that we introduced in Section 2.1.2.

**3.1.1  Water–Water HBs**

In Figure 7 and Figure 8, we have shown the CTCF and ITCF plots for water in the whole hydration shell of various DNDs that are solvated in different solutions. The former only includes results for $MgCl_2$ solution of either of DND–H, DND–NH$_2$, or DND–OH particles, as plots for their other salt solutions display almost identical behavior. However, Figure 8 reports corresponding plots of DND–COOH for all salt solutions, where we can readily observe the specific cation effect on the decay behavior of correlation functions. Corresponding plots for water in the bulk region of the neutral DND–H solutions are also shown in Figure 7 and Figure 8. We can qualitatively identify in Figure 7 and Figure 8 that the ITCF's decay rates are slower than CTCF's. The difference in decay rates is rooted in different dynamic characteristics of HBs that these time correlation functions measure. As we pointed out in Section 2.1.2, the latter considers a HB pair correlated at $t_0$ and $t_0 + t$ times, only if the donor–acceptor pair maintains its HB throughout "$t$" interval. In contrast, the former allows the HB between the pair to break and reform during time $t$.

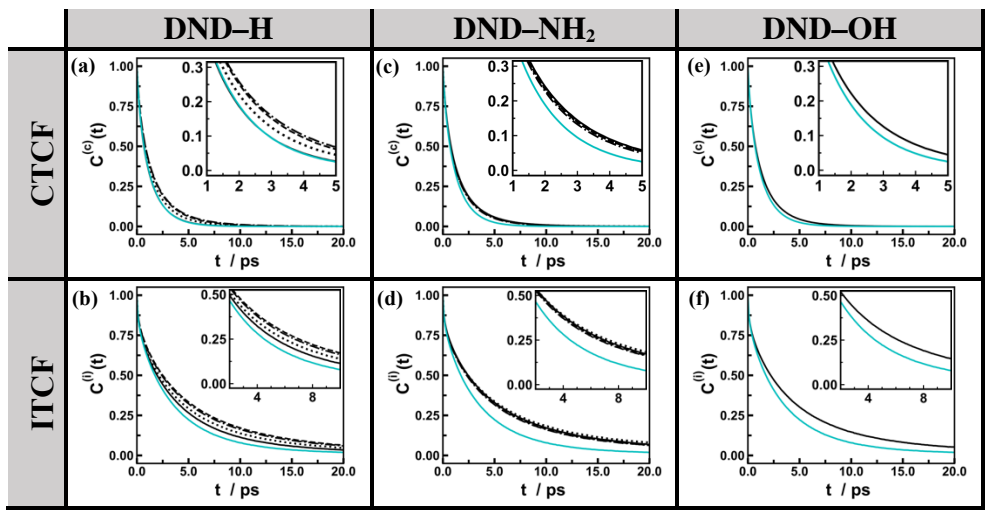

**Figure 7.** The CTCF and ITCF plots of water-water HBs in the whole hydration shell (black lines) of (a)-(b) DND–H, (c)-(d) DND–NH$_2$, (e)-(f) DND–OH, each solvated in MgCl$_2$ aqueous solution. For comparison, corresponding plots of bulk water (cyan solid line) are also shown. Black solid, dashed, dash-dotted, and dotted lines represent DNDs with 0, +28, +56, and +84 charges, respectively.

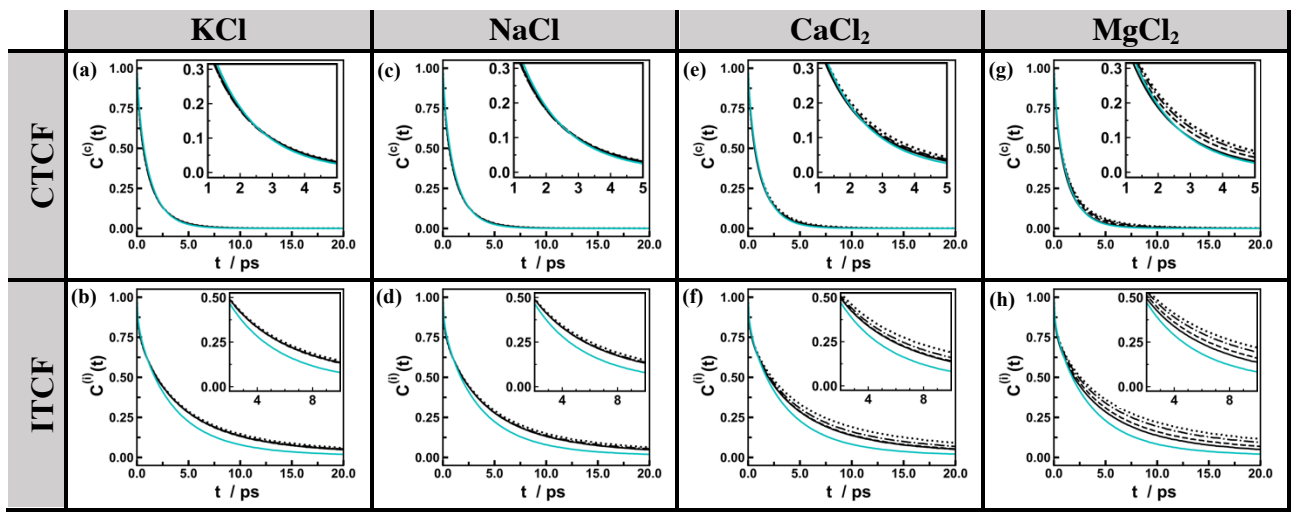

**Figure 8.** Same as Figure 7, but for DND–COOH of various charges in four different salt solutions: KCl, NaCl, CaCl$_2$, and MgCl$_2$. Black solid, dashed, dash-dotted, and dotted lines represent DNDs with 0, –28, –56, and –84 charges, respectively.

We have demonstrated the average lifetime $\tau^{(c)}$ and the structural relaxation time $\tau^{(i)}$ of HBs in Figure 9 and listed them in the SI (Table S.7 to Table S.11). We note a couple of interesting observations in Figure 9, which are discussed below.

1) Values of $\tau^{(i)}$ are larger than those of $\tau^{(c)}$, where the trend is congruent with the abovementioned ITCF's slower decay rates than CTCF's. Depending on the surface chemistry of the DND and in some cases the type of the solvated salt, the ratio of $\tau^{(i)}$ to $\tau^{(c)}$ ranges from 3 to 5.

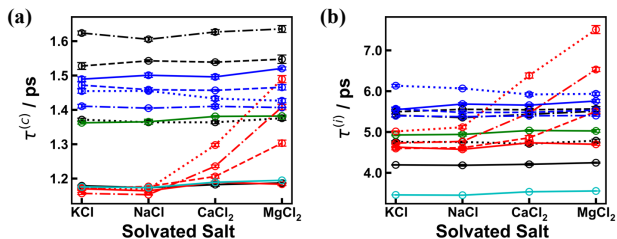

**Figure 9.** (a) The lifetime and (b) the structural relaxation time of water-water HBs in the whole hydration shell of different DNDs with various surface chemistries, solvated in different salt solutions. Line styles and colors represent the same things as those in Figure 3.

2) Among the uncharged DNDs, water-water HBs in the whole hydration layer of DND–NH$_2$ experience the longest average lifetime and also structural relaxation time. In this regard, we can sort the neutral DNDs as follows:

**DND–H ≈ DND–COOH < DND–OH < DND–NH$_2$**

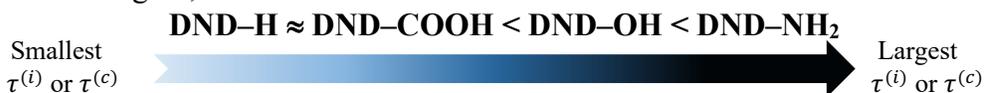

Smallest $\tau^{(i)}$ or $\tau^{(c)}$ — Largest $\tau^{(i)}$ or $\tau^{(c)}$

Interestingly, we also observed in our previous study a similar ordering for the relaxation time of the reorientational dynamics of water's OH bonds ($\tau_{corr}^{oh}$) in the first hydration layer of the neutral DNDs, although $\tau_{corr}^{oh}$ corresponding to the neutral DND–COOH was almost 30% higher than that of the neutral DND–H. This difference between the ordering of the current study and that of our previous study can be attributed to the fact that $\tau^{(i)}$ or $\tau^{(c)}$ values, as opposed to the aforementioned $\tau_{corr}^{oh}$ values, are obtained for the whole hydration shell of DNDs. Nevertheless, the overall resemblance between the two orderings suggests that there is an association between the degree to which the reorientations of water's OH bonds are constrained in the first hydration layer of the neutral DNDs and water-water HB lifetimes in DNDs' whole hydration shell.

3) We observe an interesting pattern for the case of DND–H. On the one hand, the lifetime of HBs between water molecules in the whole hydration layer of the uncharged DND–H is almost identical to that in the bulk water (see Figure 9(a)). This can be attributed to the hydrophobic nature of DND–H, where the interfacial water does not form any HBs with facets of the DND. On the other hand, the aforementioned lifetime jumps by 31% and 38% in the case of DND–H with +28 and +56 net charges, respectively. It is quite fascinating to note that DND–H with +56 and +28 charges have, respectively, the first and second largest $\tau^{(c)}$ values among all DNDs with different surface chemistries and in various salt solutions. This trend is in accord with enhancements in water-water HBs around DND–H with +56 charges that we observed earlier in Figure 4. As the DND–H obtains additional

+28 charges with respect to that with +56 charges, its surrounding water-water HB lifetime drops by %16. However, it is still longer than the bulk water-water HB lifetime. As we noted in Section S.3, we attribute this effect to disruptions in water-water HBs induced by $Cl^-$ anions, which are adsorbed onto facets of DND–H with +84 charges in higher concentrations than they do in the case of DND–H with +28 and +56 charges.

4) The specific cation effect that we noted in Figure 8 is also reflected in values of both $\tau^{(c)}$ and $\tau^{(i)}$ for water in the whole hydration layer of the negatively charged DND–COOH. While $\tau^{(c)}$ values corresponding to DND–COOH with all amounts of surface charges in NaCl and KCl solutions are equal to or smaller than those of bulk water, $CaCl_2$ and $MgCl_2$ solutions of the negatively charged DND–COOH exhibit improved $\tau^{(c)}$ values. In particular, we observe 25% and 60% increases in, respectively, $\tau^{(c)}$ and $\tau^{(i)}$ of water in the whole hydration shell of DND–COOH with –84 charges solvated in $MgCl_2$ solution with respect to those of water surrounding the neutral DND–COOH. Corresponding improvements in $CaCl_2$ solution are 8% and 35%, respectively. Thus, solutions of the charged DND–COOH containing divalent cations, particularly $Mg^{2+}$, are associated with remarkably higher correlation times than those bearing monovalent cations. This effect can be attributed to higher charge densities of $Ca^{2+}$ and $Mg^{2+}$ cations than those of $K^+$ and $Na^+$ cations, which results in impacting the HB dynamics of water in two ways. First, adsorbed $Mg^{2+}$ cation on the charged DND–COOH together with $–COO^-$ surface-bound anion significantly slow down the reorientational dynamics of water through a cooperative mechanism.[119] In other words, the former constrains the reorientations of water's dipole moment, while the latter restricts the rotational motion of OH bond of the same water by forming a HB with it. The effect of the cooperative slowdown can extend beyond the first hydration shell of strongly hydrated ions, which leads to locking in the HB network of water in multiple directions.[58,119] $Ca^{2+}$ cation also exhibits a similar, yet to a lesser degree, behavior. Second, the strength of water-water HBs in the whole hydration shell of $Mg^{2+}$ cation is significantly enhanced, compared with those around other cations and also bulk water. This observation is supported by the trends in CTCF and ITCF plots and their associated relaxation times for water in the whole hydration shell of solvated ions that are shown in Figure 10. More specifically, Figure 10(c-d) show that $\tau^{(c)}$ and $\tau^{(i)}$ values (also, listed in Table S.12 of the SI) corresponding to $Mg^{2+}$ cation are, respectively, 2 and 2.6 times as large as those of bulk water. Both of these correlation times for water around $Ca^{2+}$ cation are very close to those of bulk water, although $\tau^{(i)}$ corresponding to the former is almost 8% larger than that of the latter. In contrast, water-water HBs in the whole hydration shell of $K^+$ and $Na^+$ cations have noticeably shorter lifetimes and smaller structural relaxation times than those in the bulk region.

5) We identify in Figure 9 an anomalous behavior for water around $DND–NH_2$ having +84 charges with respect to that around the same DND but with lower charges. Both $\tau^{(c)}$ and $\tau^{(i)}$ values gradually decrease as the neutral $DND–NH_2$ obtains +28 and +56 charges. However, this trend gets reverse in the case of $DND–NH_2$ with +84 charges. Particularly, its corresponding $\tau^{(i)}$ values are almost 5% and 11% larger than those of water around the neutral $DND–NH_2$ and $DND–NH_2$ with +56 charges, respectively. We attribute this behavior to the same abovementioned cooperative slowdown of water's reorientational dynamics, yet with different players here. That is, the surface-bound $-NH_3^+$ group restricts the rotational motion of the interfacial water's dipole moment, whereas adsorbed $Cl^-$ anion constrains its OH bond's reorientations through an established HB. However, the cooperative slowdown is only effective in the case of DND–NH$_2$ with +84 charges, where there are sufficient oppositely charged ions at the interface between the DND and water.

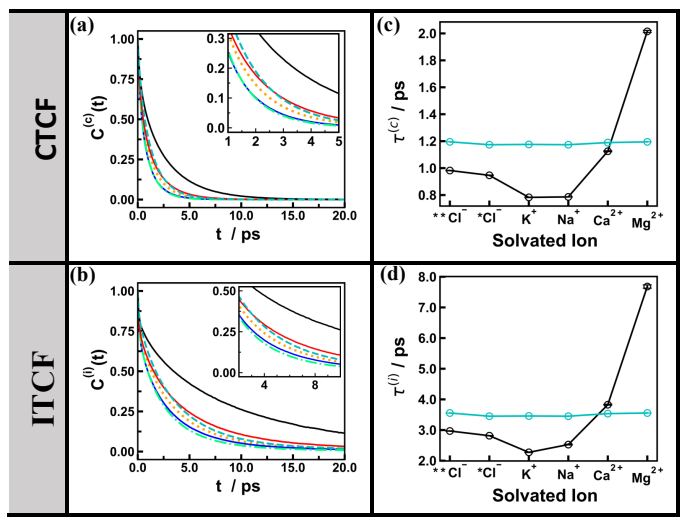

**Figure 10.** (a)-(b) CTCF and ITCF plots, (c) the lifetime, and (d) the structural relaxation time of water-water HBs in the whole hydration shell of solvated ions in four different aqueous salt solutions of the neutral DND–H. **$Cl^-$ and *$Cl^-$ refer to chloride anions in MgCl$_2$ and NaCl salts, respectively. Line styles and colors represent the same things as what described in Figure 6.

We wrap up this section by adding some noteworthy comments about the trends we observed in Figure 10(c-d). The remarkably enhanced and weakened water-water HBs around, respectively, $Mg^{2+}$ and $K^+$ cations are congruent with their strong structure-making and structure-breaking characteristics, which are well-documented in the literature.[83] On the other hand, $Ca^{2+}$ exhibits some moderate structure-making behavior, which is consistent with reported behavior in prior studies.[15,31,32,34,48–52] Although some studies[70] have categorized $Na^+$ as a structure-maker, we have found the opposite behavior for it. Our finding is based on our calculated values for the lifetime and structural relaxation time of water-water HBs around $Na^+$ cation, which are smaller than those of bulk water. Indeed, Marcus has considered $Na^+$ as a borderline case in his categorization of ions as water structure-maker or structure-breaker.[68] Compared with $K^+$, $Na^+$ imparts more order in its first hydration shell water by imposing stronger restrictions on reorientations of water's dipole moment in its immediate vicinity.[119] However, the less restricted first hydration shell water of $K^+$ forms more HBs than that of $Na^+$ with surrounding water molecules or even other polar moieties. It is consistent with our calculated number of water-water HBs around ions in the present work and also findings of previous studies.[58,75]

### 3.1.2 Water–Site HBs

Figure 11 shows CTCF and ITCF plots for HBs between the interfacial water and various oxygen atoms (introduced in the caption of Figure 11) of –COOH and –COO$^-$ surface functional groups of DND–COOH with different number of charges in MgCl$_2$ solution. The decay characteristics of CTCF plots reveal that there are primarily two types of water-site HBs for DND–COOH as follows: 1) water's oxygen (OW) accepts an HB from the hydroxyl part of –COOH group, 2) water donates a HB to the oxygen of –COO⁻ group (only for charged DND–COOH). ITCF plots in Figure 11(b, d, f, h) show that water molecules attempt to donate HBs to the doubly-bonded oxygen (O2) and the oxygen (OH) of the hydroxyl part of –COOH group. However, as we can see in corresponding CTCF plots (Figure 11(a, c, e, g)), these HBs are very unstable and quickly break.

The calculated lifetimes and structural relaxation times of water-site HBs for DND–COOH system are shown in Figure 12 and also presented in Table S.13 of the SI. Values of $\tau^{(c)}$ in Figure 12(a, c) reflect stronger water-site HBs (i.e., OW–O and OH–OW) than water-water HBs in either the whole hydration shell of DND–COOH or in the bulk region. Particularly, we note the relatively large values for $\tau^{(c)}$ and $\tau^{(i)}$ corresponding to HBs donated by –COOH groups to the interfacial water of the neutral DND–COOH. It provides an additional evidence for lower number of water-water HBs in the whole hydration shell of this DND, as we observed in Figure 3 and Figure 4. Furthermore, the HB lifetimes in Figure 12(e) agree well with the abovementioned point about the weak and unstable HBs between water and O2 atom.

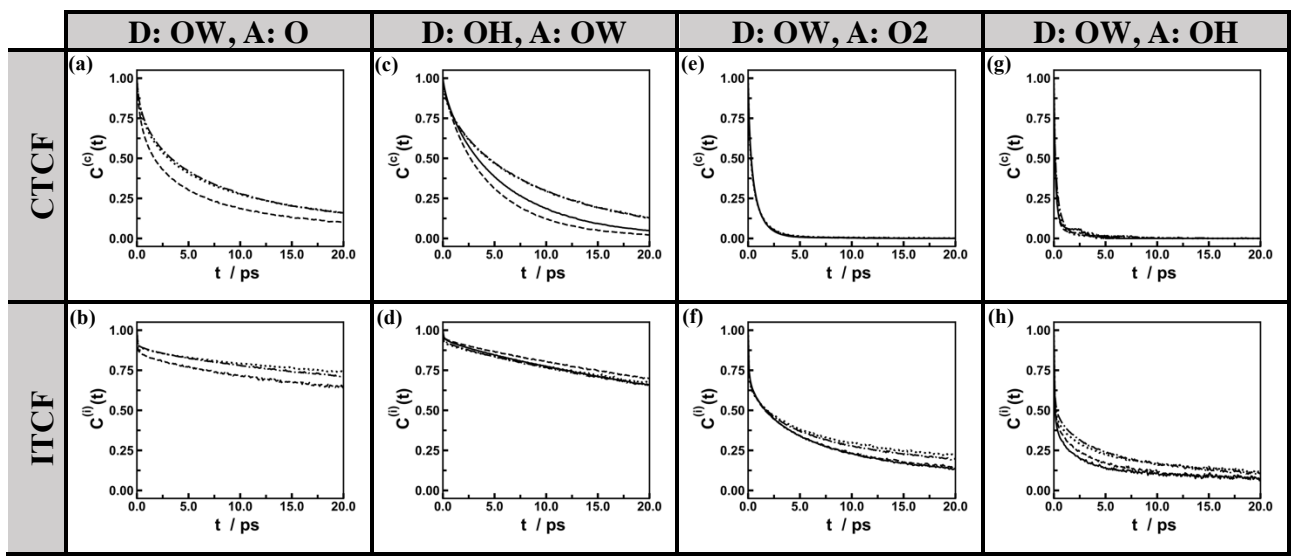

**Figure 11.** The CTCF and ITCF plots of different water-site HBs for water near DND–COOH with various charges in MgCl$_2$ aqueous solution: (a)-(b) water as HB donor to –COO⁻, (c)-(d) –COOH as HB donor to water, (e)-(f) water as HB donor to O2 of –COOH, and (g)-(h) water as HB donor to O of –COOH. The heavy atoms of HB donor (D) and HB acceptor (A) of each of these water-site HBs are introduced in the header. OW, O, O2, and OH represent, respectively, water oxygen, oxygen in –COO⁻, doubly-bonded oxygen of –COOH and oxygen in the hydroxyl part of –COOH. Solid, dashed, dash-dotted, and dotted lines represent DNDs with 0, –28, –56, and –84 charges, respectively.

An association between the type of the solvated cation and correlation times of water–COO⁻ HBs emerges in Figure 12(a-b) that piques our interest. In fact, the order in which their $\tau^{(c)}$ and $\tau^{(i)}$ both increase with regard to the type of the cation follows Na⁺ < K⁺ < Ca²⁺ < Mg²⁺, for each specific number of charges on DND–COOH. However, there are two exceptions–that is, $\tau^{(c)}$ values corresponding to DND–COOH with –84 charges in NaCl and KCl solutions and $\tau^{(i)}$ values corresponding to DND–COOH with –28 charges in KCl and CaCl$_2$ solutions are almost identical. Nonetheless, we find this ordering as similar to what we observed in our previous work[119] for the relaxation time of the reorientational dynamics of water's dipole moment in the first hydration layer of the charged DND–COOH. Thus, this indicates the role of cations in the cooperative mechanism, which we described in part 4 of Section 3.2.1, in regulating the dynamics of water–COO⁻ HBs.

We now turn our attention to water-site HB's dynamics of DND–NH$_2$. The correlation functions and their associated correlation times are shown in Figure 13 and Figure 14, respectively. We are concerned with three types of water-site HBs, denoted as $HB_{WN}^{(1)}$, $HB_{WN}^{(2)}$, and $HB_{WN}^{(3)}$. The first one is related to HBs donated by $-NH_3^+$ groups on the charged DND to the nearby water, whereas the second and third ones represent HBs of water, respectively, donated to and accepted from –NH$_2$ groups. The numerical values for the corresponding $\tau^{(c)}$ and $\tau^{(i)}$ are presented in Table S.14 of the SI.

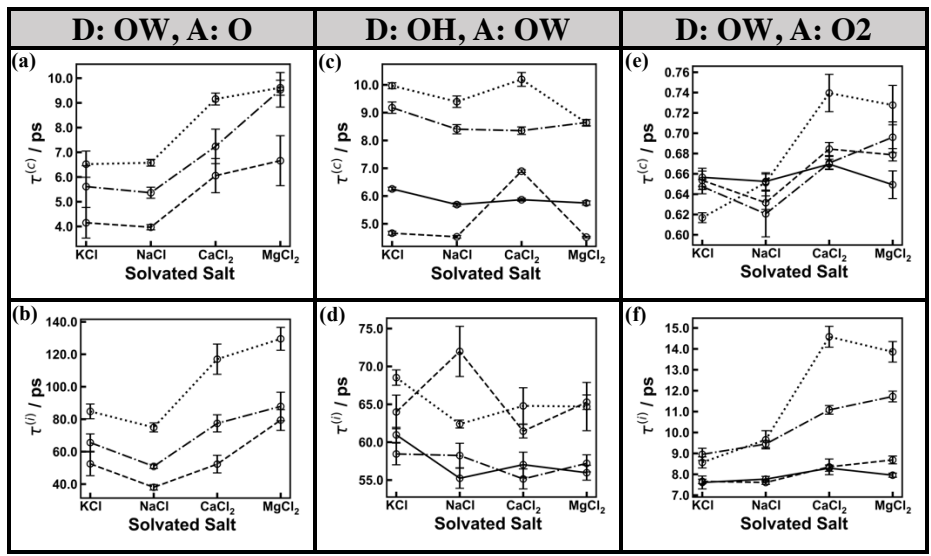

**Figure 12.** The lifetime ($\tau^{(c)}$) and the structural relaxation time ($\tau^{(i)}$) of different water-site HBs (as introduced in Figure 11) for water near DND–COOH with various charges in different aqueous salt solutions. Solid, dashed, dash-dotted, and dotted lines represent DNDs with 0, –28, –56, and –84 charges, respectively.

Some interesting patterns emerge in the lifetime of the aforementioned HBs, as the neutral DND–NH$_2$ acquires increasing number of positive charges via the protonation of its –NH$_2$ groups (see Figure 14(a, c, e)). While $\tau^{(c)}$ values of $HB_{WN}^{(1)}$ and $HB_{WN}^{(3)}$ increase, those of $HB_{WN}^{(2)}$ decrease with only one exception–that is, $\tau^{(c)}$ values of $HB_{WN}^{(2)}$ for DND–NH$_2$ with +84 charges are larger than those of DND–NH$_2$ with +56 charges. Nonetheless, we infer that $HB_{WN}^{(2)}$ is the dominant type of water-site HB for the neutral DND–NH$_2$. In other words, –NH$_2$ groups on this DND primarily act as HB acceptors from the interfacial water. However, as the number of $-NH_3^+$ groups on the DND increases, –NH$_2$ groups gradually switch their role from being a primarily HB acceptor to an HB donor. We ascribe this effect to the change induced by $-NH_3^+$ groups in the preferential orientation of the first hydration layer water of the charged DND–NH$_2$, which we described in Section S.3.

We note relatively high values of $\tau^{(c)}$ and $\tau^{(i)}$ (on average, around 4.5 and 105 ps respectively) for $HB_{WN}^{(1)}$ of DND–NH$_2$ with +84 charges in Figure 14(a-b), which are approximately 4 and 30 times larger than those of bulk water-water HBs. It implies that the HB of a specific water$-NH_3^+$ pair breaks, on average, after 4.5 ps of its formation. However, the water sojourns nearby its partner (i.e., $-NH_3^+$) for approximately 105 ps during which their initial HB undergoes intermittent reformation and breakage. Thus, the water molecule's oxygen atom and its dipole moment are remarkably restricted in the immediate vicinity of DND–NH$_2$ with +84 charges. This observation corroborates the relatively long mean residence time[58] and the relaxation time of the dipolar reorientational dynamics[119] of water in the closest water layer to DND–NH$_2$ with +84 charges.

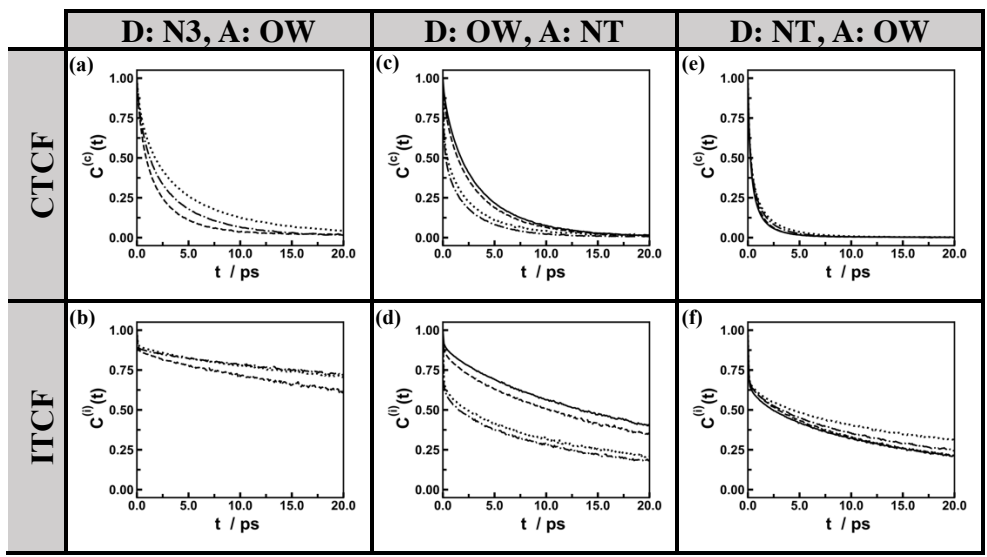

**Figure 13.** Same as Figure 11, but for water near DND–NH$_2$: (a)-(b) water as HB acceptor from $-NH_3^+$ ($HB_{WN}^{(1)}$ type), (c)-(d) water as HB donor to –NH$_2$ ($HB_{WN}^{(2)}$ type), and (e)-(f) water as HB acceptor from –NH$_2$ ($HB_{WN}^{(3)}$ type). Heavy atoms of HB donor (D) and HB acceptor (A) of each of these water-site HBs, as introduced in the header, are: OW: water oxygen, N3: N atom in $-NH_3^+$, and NT: N atom in –NH$_2$. Solid, dashed, dash-dotted, and dotted lines represent DNDs with 0, +28, +56, and +84 charges, respectively.

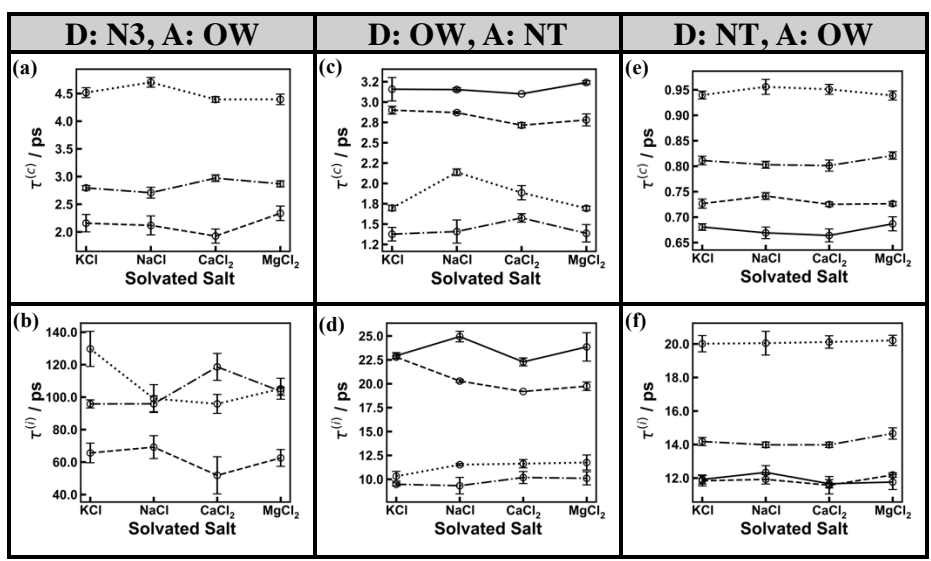

**Figure 14.** The lifetime ($\tau^{(c)}$) and the structural relaxation time ($\tau^{(i)}$) of different water-site HBs (as introduced in Figure 13) for water near DND–NH$_2$ with various charges in different aqueous salt solutions. Solid, dashed, dash-dotted, and dotted lines represent DNDs with 0, +28, +56, and +84 charges, respectively.

We wrap up this section by looking into the dynamics of HBs between water and –OH groups on DND–OH. Figure 15 demonstrates CTCF and ITCF plots of these HBs as well as their associated correlation times. There exist two types of water-site HBs at the interface with DND–OH, referred to as $HB_{WOH}^{(1)}$ and $HB_{WOH}^{(2)}$. The former is formed as a result of the hydrogen donation by water to –OH group, whereas the reverse takes place to form the latter. Our previous investigations on the preferential orientation of the interfacial water showed to us that $HB_{WOH}^{(2)}$ are more prevalent on {111} facets of DND–OH than on its {100} facets.[58]

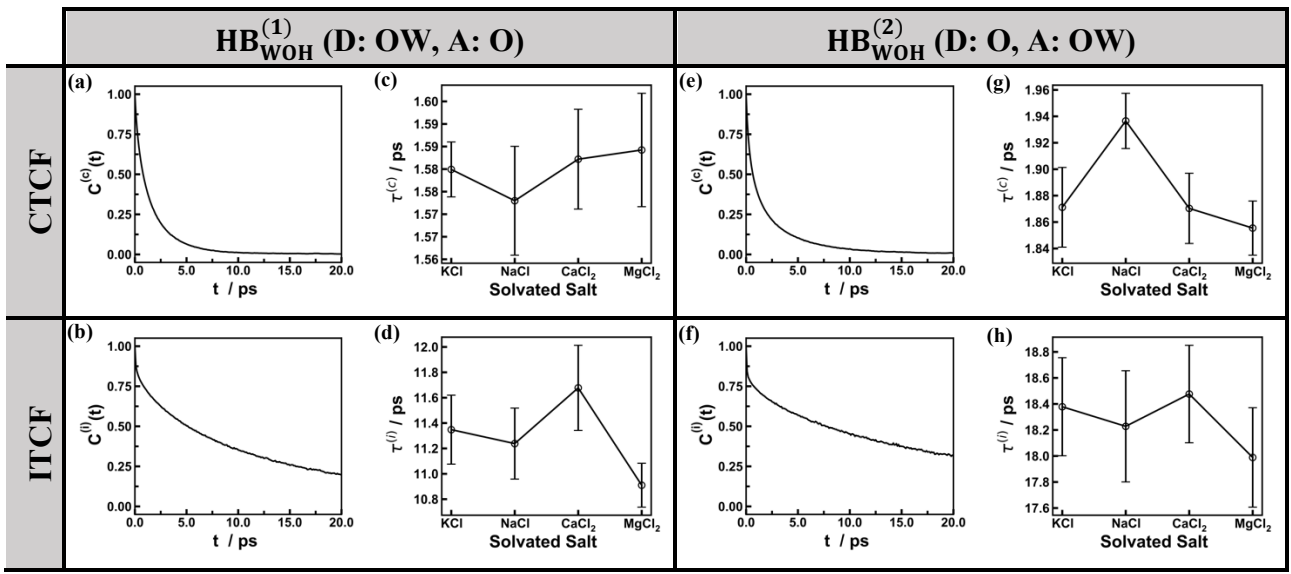

**Figure 15.** The CTCF and ITCF plots and their related correlation times (i.e., lifetime $\tau^{(c)}$ and structural relaxation time $\tau^{(i)}$) of different water-site HBs for water near DND–OH: (a)-(d) water as HB donor to –OH ($HB_{WOH}^{(1)}$ type), (e)-(h) water as HB acceptor from –OH ($HB_{WOH}^{(2)}$ type). Heavy atoms of HB donor (D) and HB acceptor (A) of each of these water-site HBs, as introduced in the header, are: OW: water oxygen, O: in –OH. Plots in (a)-(b) and (e)-(f) correspond to $MgCl_2$ aqueous solution of DND–OH.

Despite their narrow ranges of values, $\tau^{(c)}$ and $\tau^{(i)}$ of both $HB_{WOH}^{(1)}$ and $HB_{WOH}^{(2)}$ exhibit some noteworthy patterns in Figure 15 (see Table S.15 in the SI for numerical values). First of all, both $\tau^{(c)}$ and $\tau^{(i)}$ values of $HB_{WOH}^{(2)}$ appear to be larger than those of $HB_{WOH}^{(1)}$ by approximately 0.3 and 7.0 ps, respectively. Second, we identify $Na^+ < K^+ < Ca^{2+} < Mg^{2+}$ ordering for solvated cations with respect to which $\tau^{(c)}$ values of $HB_{WOH}^{(1)}$ appear to increase (see Figure 15(c)). Interestingly, it is the same ordering we observed for $\tau^{(c)}$ of water–$COO^-$ HBs in Figure 12(a). Conversely, the reverse ordering emerges in Figure 15(g) for $\tau^{(c)}$ values of $HB_{WOH}^{(2)}$. Just as water–$COO^-$ HBs, the cooperative mechanism can also explain the aforementioned ordering for $HB_{WOH}^{(1)}$, although its effect on the latter is much less pronounced than that on the former. This mechanism particularly becomes pertinent here, when we take into account the adsorption behavior of cations on DND–OH that are reported elsewhere.[119] In other words, cations adsorb more predominantly onto {100} facets of this DND than they do on its {111} facets, where the monovalent and divalent cations form, respectively, CIP and SIP associations with –OH groups. In contrast, $Cl^-$ anions primarily adsorb on {111} facets of DND–OH.

### 3.1.3 Water–Anion HBs

Figure 16 demonstrates plots of CTCF and ITCF along with their associated correlation times for HBs between water molecules and $Cl^-$ in different salt solutions of various DNDs.

Corresponding correlation times of bulk water in four different salt solutions of the neutral DND–H are also shown in Figure 16(e-f) for comparison. First of all, we observe that $\tau^{(c)}$ and $\tau^{(i)}$ of water–Cl$^-$ HBs (which are listed in the SI, Table S.16 to Table S.19) are larger than those of bulk water by, on average, 1.8 and 4 factors, respectively. Differences between monovalent cation–chloride and divalent cation–chloride solutions arise from different forcefield parameters used for Cl$^-$ anions in the respective salts. Second, lifetimes and structural relaxation times of water–Cl$^-$ HBs in solutions of DND–H and DND–NH$_2$ with +56 and +84 charges markedly differ from those of other DND solutions (see Figure 16(e-f)). We attribute this effect to an interplay between positive charges of aforementioned DNDs and adsorbed Cl$^-$ anions onto their surfaces. On the one hand, DND–H and DND–NH$_2$ with +56 and especially +84 charges adsorb higher concentrations of Cl$^-$ anions than other DNDs do. On the other hand, as we found in our previous study,[119] the concentrated Cl$^-$ anions markedly slow down the reorientation of their hydrating water's OH bond, which subsequently hinders water–Cl$^-$ HB breakage. It corroborates the results reported by Stirnemann *et al.*, where they observed weakly hydrated ions, in our case Cl$^-$, at high concentrations significantly retard the dynamics of their nearby water.[83]

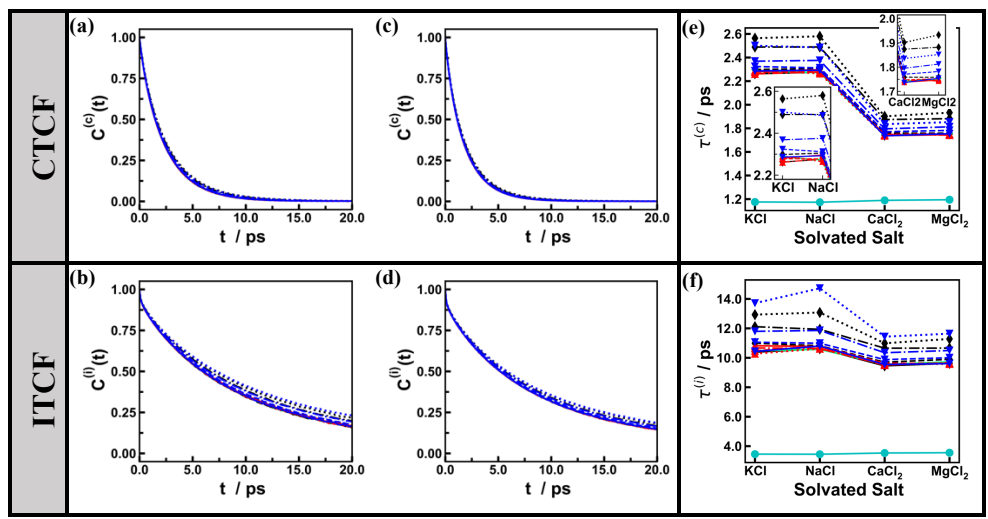

**Figure 16.** The CTCF and ITCF plots and their related correlation times (i.e., lifetime $\tau^{(c)}$ and structural relaxation time $\tau^{(i)}$) of water–Cl$^-$ HB in different chloride salt solutions of various DNDs. Plots in (a)-(b) and (c)-(d) correspond to NaCl and MgCl$_2$ aqueous solutions of various DNDs. Line styles and colors are the same as those in Figure 3.

## 4 Conclusion

Both the average number of HBs per water and their dynamic characteristics in the hydration shell of various DNDs and solvated ions in the aqueous solution have been determined through MD simulations. Our main focus has been on obtaining a microscopic picture of how different factors impact the HBs of water near a single DND particle, which has important implications for its fluorescent property and colloidal stability.[15,31,32,34,48–52] The studied factors include hydrophobic vs. hydrophilic and uncharged vs. charged surface chemistries of the DND on the one hand and solvated ions of differing charge densities on the other hand.

We have found –NH$_2$ group on the neutral DND–NH$_2$ as a strong HB acceptor from water, whereas –COOH group on the neutral DND–COOH appears as a strong HB donor to water. However, –OH group on DND–OH exhibit both HB donor and acceptor tendencies toward water, although the former result in relatively stronger HBs with water and is more prevalent on {111} facets.[58] Nevertheless, both kinds of HBs established by –OH groups are weaker than the aforementioned HBs created by –$NH_2$ and –COOH groups. On the other hand, the hydrophobic neutral DND–H promotes some dangling water's OH in the immediate vicinity of its facets. Thus, these different DND-water interactions induce distinct preferential orientations[58] in the interfacial water that leads to the formation of water-water HBs with differing strengths. Indeed, we have found that the average lifetime and hence the strength of water-water HBs in the whole hydration shell of neutral DNDs and that of bulk water grows according to bulk water ≈ DND–H ≈ DND–COOH < DND–OH < DND–$NH_2$ ordering in all different salt solutions. The observed ordering for the structural relaxation time of aforementioned HBs is bulk water < DND–H < DND–COOH < DND–OH < DND–$NH_2$. Interestingly, the average number of HBs of water with its neighboring water molecules in the bulk region and also near neutral DNDs increase in the order of DND–COOH < DND–OH < DND–$NH_2$ < DND–H < bulk water.

The abovementioned behavior changes in certain ways with the addition of some charges to DNDs. Strikingly, we have observed the longest water-water HB's lifetime (hence, the strongest water-water HB) in the whole hydration shell of DND–H with +56 charges, compared with bulk water and also with that of DNDs having different surface chemistries in various salt solutions. We ascribe this observation to the promotion of water-water HBs in the closest layer of water to facets of DND–H with +56 charges. It results from the repulsion of water's OH bonds from the DND's positively charged facets that would otherwise have been dangling on facets of the natural DND–H. The reduction of dangling water's OH population in the vicinity of positively charged hydrophobes has also been reported elsewhere[86]. However, the water-water HB lifetime drops by the addition of +28 charges to DND–H with +56 charges. We attribute this effect to two factors. First, the positively charged facets of the resulting DND (i.e., DND–H with +84 charges) push the immediate nearby water's OH bonds too far away from its facets. Second, DND–H with +84 charges adsorbs more $Cl^-$ anions, which are water-water structure-breaker, to its facets than DND–H with lower charges does. Thus, water-water HB is not promoted as much as what we mentioned in the case of DND–H with +56 charges. This disruption in water-water HBs around DND–H with +84 charges might be related to the formation of micro-sized aggregates of DND–H with positive zeta potential in aqueous solution containing 10-20 mM HCl acid.[120]

The effect of the charged DND–COOH, whose surface is covered by certain amounts of –$COO^-$ groups, on the interfacial water's HBs depends on the charge density of adsorbed cations. Upon the adsorption of either of $K^+$ and $Na^+$ cations on the charged DND–COOH, the DND's nearby water-water HBs have the shortest lifetime compared with all other DNDs of different surface chemistries. Their lifetimes are even slightly shorter than that of bulk water and also that of the neutral DND–COOH. These effects, to some extent, can be explained in terms of the structure-breaking nature of these cations, which we have explained before. However, strongly hydrated $Ca^{2+}$ and more prominently $Mg^{2+}$, whose charge densities are both higher than that of either $K^+$ or $Na^+$, appear as game-changers. In other words, water-water HB's average lifetime near the charged DND–COOH in either $CaCl_2$ or $MgCl_2$ solutions increases with the increasing number of –$COO^-$ groups on the DND's surface. In the extreme case of 84 –$COO^-$ groups on DND–COOH in $MgCl_2$ solution, the water-water HB lifetime tends to that around the neutral DND–$NH_2$. Furthermore, the multiply charged cations, especially $Mg^{2+}$, are associated with the longest lifetime for water–$COO^-$ HB for the case of DND–COOH with –84 charges, compared with all other types of HBs in this study.

The abovementioned effects of $Ca^{2+}$ and $Mg^{2+}$ can be explained in terms of the cooperative hydration mechanism, implemented by either of these cations and the surface-bound –$COO^-$ anion. We proposed this mechanism in our previous work[119], where we investigated the translational and orientational dynamics of water around the same DNDs as those of the current study. In particular, our previous study[119] revealed the significant influence of $Ca^{2+}$ and $Mg^{2+}$ cations on slowing down the translational and orientational dynamics of the interfacial water of DND–COOH with –84 charges. The present study also shows that the adsorption of either $Ca^{2+}$ or $Mg^{2+}$ (particularly the latter) on facets of DND–COOH with –84 charges leads to the longest structural relaxation time of water-water HBs in the whole hydration shell of this DND, in comparison to all other DNDs and bulk water. Thus, these two observations together point to the collective reorganization of water's HB network that results from the coupled translational-rotational motions of its constituent molecules.[96–100,119]

The addition of positive charges (through $-NH_3^+$ surface groups) to the neutral DND–$NH_2$ gradually causes the DND to switch from a predominantly HB acceptor from water to a HB donor to water. In the extreme case of DND with +84 charges, we have observed relatively high lifetimes and structural relaxation times for both water–water and water–$NH_3^+$ HBs. We explain this effect in terms of a similar cooperative hydration mechanism[119] that we mentioned above but implemented here by $Cl^-$ anion and surface-bound $-NH_3^+$ cation.

In summary, we have been able to obtain a molecular-level understanding of how various surface chemistries of DNDs and different solvated ions impact HBs of water with other water molecules and also with surface functional groups of DNDs. We estimated the strength of HBs and environmental constraints on their restructuring by calculating the lifetime and structural relaxation time of HBs.[95,96] However, it would be beneficial to carry out further investigations to quantitatively assess the enthalpic and entropic contributions of aforementioned factors into HBs of DNDs' interfacial water.[23]

**References**

1. Bagchi, B. Water solvation dynamics in the bulk and in the hydration layer of proteins and self-assemblies. *Annu. Rep. Prog. Chem., Sect. C* vol. 99 127–175 (2003).
2. Nakagawa, H., Joti, Y., Kitao, A. & Kataoka, M. Hydration affects both harmonic and anharmonic nature of protein dynamics. *Biophys. J.* **95**, 2916–2923 (2008).
3. Aggarwal, L. & Biswas, P. Hydration Water Distribution around Intrinsically Disordered Proteins. *J. Phys. Chem. B* **122**, 4206–4218 (2018).
4. Mukherjee, K., Schwaab, G. & Havenith, M. Cation-specific interactions of protein surface charges in dilute aqueous salt solutions: A combined study using dielectric relaxation spectroscopy and Raman spectroscopy. *Phys.Chem.Chem.Phys.* **20**, 29306–29313 (2018).
5. Siebert, T., Guchhait, B., Liu, Y., Costard, R. & Elsaesser, T. Anharmonic Backbone Vibrations in Ultrafast Processes at the DNA-Water Interface. *J. Phys. Chem. B* **119**, 9670–9677 (2015).
6. Liu, Y., Guchhait, B., Siebert, T., Fingerhut, B. P. & Elsaesser, T. Molecular couplings and energy exchange between DNA and water mapped by femtosecond infrared spectroscopy of backbone vibrations. *Struct. Dyn. 4* **4**, (2017).
7. Alizadeh, M., Azar, P. A., Mozaffari, S. A., Karimi-Maleh, H. & Tamaddon, A. M. A DNA Based Biosensor Amplified With ZIF-8/Ionic Liquid Composite for Determination of Mitoxantrone Anticancer Drug: An Experimental/Docking Investigation. *Front. Chem.* **8**, 1–10 (2020).
8. Karimi-Maleh, H. *et al.* Guanine-Based DNA Biosensor Amplified with Pt/SWCNTs Nanocomposite as Analytical Tool for Nanomolar Determination of Daunorubicin as an Anticancer Drug: A Docking/Experimental Investigation. *Ind. Eng. Chem. Res.* **60**, 816–

823 (2021).
9. Su, L., Krim, J. & Brenner, D. W. Dynamics of Neutral and Charged Nanodiamonds in Aqueous Media Confined between Gold Surfaces under Normal and Shear Loading. *ACS Omega* **5**, 10349–10358 (2020).
10. Keshri, S. & Tembe, B. L. Thermodynamics of hydration of fullerols [C60(OH)n] and hydrogen bond dynamics in their hydration shells. *J. Chem. Phys* **146**, 74501 (2017).
11. Zobel, M., Neder, R. B. & Kimber, S. A. J. Universal solvent restructuring induced by colloidal nanoparticles. *Science* vol. 347 (2015).
12. Thomä, S. L. J., Krauss, S. W., Eckardt, M., Chater, P. & Zobel, M. Atomic insight into hydration shells around facetted nanoparticles. *Nat. Commun.* **10**, (2019).
13. Zobel, M. Observing structural reorientations at solvent-nanoparticle interfaces by X-ray diffraction - Putting water in the spotlight. *Acta Cryst.* **A72**, 621–631 (2016).
14. Collins, K. D. Ions from the Hofmeister series and osmolytes: Effects on proteins in solution and in the crystallization process. *Methods* **34**, 300–311 (2004).
15. Dolenko, T. A., Burikov, S. A., Rosenholm, J. M., Shenderova, O. A. & Vlasov, I. I. Diamond−Water Coupling Effects in Raman and Photoluminescence Spectra of Nanodiamond Colloidal Suspensions. *J.Phys.Chem.C* **116**, 24314–24319 (2012).
16. Fogarty, A. C., Duboué-Dijon, E., Sterpone, F., Hynes, J. T. & Laage, D. Biomolecular hydration dynamics: A jump model perspective. *Chem. Soc. Rev.* **42**, 5672–5683 (2013).
17. Adam, S. & Bondar, A.-N. Mechanism by which water and protein electrostatic interactions control proton transfer at the active site of channelrhodopsin. *PLoS One* **13**, (2018).
18. Laage, D., Elsaesser, T. & Hynes, J. T. Water Dynamics in the Hydration Shells of Biomolecules. *Chem.Rev.* **117**, 10694–10725 (2017).
19. Chen, Y., Okur, H. I., Liang, C. & Roke, S. Orientational ordering of water in extended hydration shells of cations is ion-specific and is correlated directly with viscosity and hydration free energy. *Phys. Chem. Chem. Phys.* **19**, 24678–24688 (2017).
20. Ding, Y., Hassanali, A. A., Parrinello, M., Designed, M. P. & Per-Formed, A. A. H. Anomalous water diffusion in salt solutions. *PNAS* **111**, (2014).
21. Bergqvist, S., Williams, M. A., O'Brien, R. & Ladbury, J. E. Heat Capacity Effects of Water Molecules and Ions at a Protein-DNA Interface. *J. Mol. Biol.* **336**, 829–842 (2004).
22. Celik, Y. *et al.* Microfluidic experiments reveal that antifreeze proteins bound to ice crystals suffice to prevent their growth. *Proc. Natl. Acad. Sci. U. S. A.* **110**, 1309–1314 (2013).
23. Duboué-Dijon, E. & Laage, D. Comparative study of hydration shell dynamics around a hyperactive antifreeze protein and around ubiquitin. *J. Chem. Phys.* **141**, 22–529 (2014).
24. Fogarty, A. C. & Laage, D. Water Dynamics in Protein Hydration Shells: The Molecular Origins of the Dynamical Perturbation. *J. Phys. Chem. B* **118**, 53 (2014).
25. Zhang, Y. *et al.* Effects of ionic hydration and hydrogen bonding on flow resistance of ionic aqueous solutions confined in molybdenum disulfide nanoslits: Insights from molecular dynamics simulations. *Fluid Phase Equilib.* **489**, 23–29 (2019).
26. Mu, L., Shi, Y., Hua, J., Zhuang, W. & Zhu, J. Engineering Hydrogen Bonding Interaction and Charge Separation in Bio-Polymers for Green Lubrication. *J. Phys. Chem. B* **121**, 5669–5678 (2017).
27. Levy, Y. & Onuchic, J. N. Water Mediation in ProteinFolding and MolecularRecognition. *Annu. Rev. Biophys. Biomol. Struct.* **35**, 389–415 (2006).
28. Mensch, C., Bultinck, P. & Johannessen, C. The effect of protein backbone hydration on the amide vibrations in Raman and Raman optical activity spectra. *Phys. Chem. Chem.*

*Phys.* **21**, 1988–2005 (2019).
29. Bellissent-Funel, M. C. *et al.* Water Determines the Structure and Dynamics of Proteins. *Chem. Rev.* vol. 116 7673–7697 (2016).
30. Yoosefian, M., Karimi-Maleh, H. & Sanati, A. L. A theoretical study of solvent effects on the characteristics of the intramolecular hydrogen bond in Droxidopa. *J. Chem. Sci.* **127**, 1007–1013 (2015).
31. Petit, T. *et al.* Unusual Water Hydrogen Bond Network around Hydrogenated Nanodiamonds. *J. Phys. Chem. C* **121**, 5185–5194 (2017).
32. Petit, T. *et al.* Probing Interfacial Water on Nanodiamonds in Colloidal Dispersion. *J. Phys. Chem. Lett.* **6**, 2909–2912 (2015).
33. Petit, T. *et al.* Surface transfer doping can mediate both colloidal stability and self-assembly of nanodiamonds. *Nanoscale* **5**, 8958–8962 (2013).
34. Vervald, A. M., Burikov, S. A., Vlasov, I. I., Shenderova, O. A. & Dolenko, T. A. Interactions of nanodiamonds and surfactants in aqueous suspensions. *Nanosyst. Phys. Chem. Math.* **9**, 49–51 (2018).
35. Laptinskiy, K. A. *et al.* Adsorption of DNA Nitrogenous Bases on Nanodiamond Particles: Theory and Experiment. *J.Phys.Chem.C* **122**, 11066–11075 (2018).
36. Petit, T., Pflüger, M., Tolksdorf, D., Xiao, J. & Aziz, E. F. Valence holes observed in nanodiamonds dispersed in water. *Nanoscale* **7**, 2987–2991 (2015).
37. Sreenivasan, V. K. A., Zvyagin, A. V. & Goldys, E. M. Luminescent nanoparticles and their applications in the life sciences. *J. Phys. Condens. Matter* **25**, (2013).
38. Lanin, S. N., Platonova, S. A., Vinogradov, A. E., Lanina, S. & Nesterenko, P. N. Regularities of adsorption of water-soluble vitamins on the surface of microdispersed sintered detonation nanodiamond. *Adsorption* **24**, 637–645 (2018).
39. Whitlow, J., Pacelli, S. & Paul, A. Multifunctional nanodiamonds in regenerative medicine: Recent advances and future directions. *J. Control. Release* **261**, 62–86 (2017).
40. Nunn, N. *et al.* Fluorescent single-digit detonation nanodiamond for biomedical applications. *Methods Appl. Fluoresc.* **6**, (2018).
41. Satar, R. *et al.* Role of Nanodiamonds in Drug Delivery and Stem Cell Therapy. *Iran. J. Biotechnol.* **14**, 130–141 (2017).
42. Siafaka, P. I., Üstündağ Okur, N., Karavas, E. & Bikiaris, D. N. Surface modified multifunctional and stimuli responsive nanoparticles for drug targeting: Current status and uses. *Int. J. Mol. Sci.* **17**, (2016).
43. Lai, L. & Barnard, A. S. Functionalized Nanodiamonds for Biological and Medical Applications. *J. Nanosci. Nanotechnol.* **15**, 989–999 (2015).
44. Chatterjee, A. *et al.* Antibacterial effect of ultrafine nanodiamond against gram-negative bacteria Escherichia coli . *J. Biomed. Opt.* **20**, 051014 (2014).
45. Pichot, V. *et al.* An efficient purification method for detonation nanodiamonds. *Diam. Relat. Mater.* **17**, 13–22 (2008).
46. Schmidlin, L. *et al.* Identification, quantification and modification of detonation nanodiamond functional groups. *Diam. Relat. Mater.* **22**, 113–117 (2012).
47. Pentecost, A., Gour, S., Mochalin, V., Knoke, I. & Gogotsi, Y. Deaggregation of Nanodiamond Powders Using Salt-and Sugar-Assisted Milling. *ACS Appl. Mater. Interfaces* **2**, undefined- (2010).
48. Vervald, A. M. *et al.* Relationship Between Fluorescent and Vibronic Properties of Detonation Nanodiamonds and Strength of Hydrogen Bonds in Suspensions. *J. Phys. Chem.*

*C* **120**, 19375–19383 (2016).
49. Burikov, S. A. *et al.* Influence of hydrogen bonds on the colloidal and fluorescent properties of detonation nanodiamonds in water, methanol and ethanol. *Fullerenes Nanotub. Carbon Nanostructures* **25**, 602–606 (2017).
50. Dolenko, T. *et al.* Evidence of carbon nanoparticle–solvent molecule interactions in Raman and fluorescence spectra. *Phys. Status Solidi A* **212**, 2512–2518 (2015).
51. Laptinskiy, K. A., Burikov, S. A., Dolenko, S. A., Shenderova, O. A. & Dolenko, T. A. Electronic effects on the interfaces "nanodiamond - surface groups–water molecules". *Fullerenes Nanotub. Carbon Nanostructures* **28**, 262–266 (2020).
52. Santacruz-Gomez, K. *et al.* Antioxidant activity of hydrated carboxylated nanodiamonds and its influence on water γ-radiolysis. *Nanotechnology* **29**, (2018).
53. Chaux-Jukic, I. *et al.* Revisiting Water Molecule Interactions with Hydrogenated Nanodiamonds: Towards Their Direct Quantification in Aqueous Suspensions. *Am J Nanotechnol Nanomed* **2**, 14–022 (2019).
54. Stehlik, S. *et al.* Water interaction with hydrogenated and oxidized detonation nanodiamonds - Microscopic and spectroscopic analyses. *Diam. Relat. Mater.* **63**, 97–102 (2016).
55. Guo, Y., Li, S., Li, W., Moosa, B. & Khashab, N. M. The Hofmeister effect on nanodiamonds: How addition of ions provides superior drug loading platforms. *Biomater. Sci.* **2**, 84–88 (2014).
56. Huang, H., Pierstorff, E., Osawa, E. & Ho, D. Active nanodiamond hydrogels for chemotherapeutic delivery. *Nano Lett.* **7**, 3305–3314 (2007).
57. Zhu, Y. *et al.* Excessive sodium ions delivered into cells by nanodiamonds: Implications for tumor therapy. *Small* **8**, 1771–1779 (2012).
58. Saberi-Movahed, F. & Brenner, D. W. What drives adsorption of ions on surface of nanodiamonds in aqueous solutions? *arXiv:2102.09187 [physics.comp-ph]* (2021).
59. Hribar, B., Southall, N. T., Vlachy, V. & Dill, K. A. How Ions Affect the Structure of Water. *J Am Chem Soc.* **124**, 12302–12311 (2002).
60. Deyerle, B. A. & Zhang, Y. Effects of hofmeister anions on the aggregation behavior of PEO-PPO-PEO triblock copolymers. *Langmuir* **27**, 9203–9210 (2011).
61. Collins, K. D. Ion hydration: Implications for cellular function, polyelectrolytes, and protein crystallization. *Biophys. Chem.* **119**, 271–281 (2006).
62. Kim, J. S., Wu, Z., Morrow, A. R., Yethiraj, A. & Yethiraj, A. Self-diffusion and viscosity in electrolyte solutions. *J. Phys. Chem. B* **116**, 12007–12013 (2012).
63. Song, J., Franck, J., Pincus, P., Kim, M. W. & Han, S. Specific ions modulate diffusion dynamics of hydration water on lipid membrane surfaces. *J. Am. Chem. Soc.* **136**, 2642–2649 (2014).
64. Choppin, G. R. & Buijs, K. Near-Infrared Studies of the Structure of Water. II. Ionic Solutions. *J. Chem. Phys.* **39**, 2042–2050 (1963).
65. Zajforoushan Moghaddam, S. & Thormann, E. The Hofmeister series: Specific ion effects in aqueous polymer solutions. *J. Colloid Interface Sci.* **555**, 615–635 (2019).
66. Kaulen, C. & Simon, U. Ion specific effects on the immobilisation of charged gold nanoparticles on metal surfaces. *RSC Adv.* **8**, 1717–1724 (2018).
67. López-León, T., Jódar-Reyes, A. B., Ortega-Vinuesa, J. L. & Bastos-González, D. Hofmeister effects on the colloidal stability of an IgG-coated polystyrene latex. *J. Colloid Interface Sci.* **284**, 139–148 (2005).

68. Marcus, Y. Effect of ions on the structure of water: Structure making and breaking. *Chem. Rev.* **109**, 1346–1370 (2009).
69. Fournier, J. A., Carpenter, W., De Marco, L. & Tokmakoff, A. Interplay of Ion-Water and Water-Water Interactions within the Hydration Shells of Nitrate and Carbonate Directly Probed with 2D IR Spectroscopy. *J. Am. Chem. Soc.* **138**, 9634–9645 (2016).
70. Collins, K. D. Charge Density-Dependent Strength of Hydration and Biological Structure. *Biophys. J.* **72**, 65–76 (1997).
71. Samoilov, Y. A new approach to the study of hydration of ions in aqueous solutions. *Discuss. Faraday Soc.* **24**, 141–146 (1957).
72. Bruni, F., Imberti, S., Mancinelli, R. & Ricci, M. A. Aqueous solutions of divalent chlorides: Ions hydration shell and water structure. *J. Chem. Phys.* **136**, (2012).
73. Collins, K. D. The behavior of ions in water is controlled by their water affinity. *Q. Rev. Biophys.* **52**, e11 (2019).
74. Shattuck, J., Shah, P., Erramilli, S. & Ziegler, L. D. Structure Making and Breaking Effects of Cations in Aqueous Solution: Nitrous Oxide Pump−Probe Measurements. *J. Phys. Chem. B* **120**, (2016).
75. Mancinelli, R., Botti, A., Bruni, F., Ricci, M. A. & Soper, A. K. Hydration of sodium, potassium, and chloride ions in solution and the concept of structure maker/breaker. *J. Phys. Chem. B* **111**, 13570–13577 (2007).
76. Gallo, P., Corradini, D. & Rovere, M. Ion hydration and structural properties of water in aqueous solutions at normal and supercooled conditions: A test of the structure making and breaking concept. *Phys. Chem. Chem. Phys.* **13**, 19814–19822 (2011).
77. Omta, A. W., Kropman, M. F., Woutersen, S. & Bakker, H. J. Negligible Effect of Ions on the Hydrogen-Bond Structure in Liquid Water. *Science* vol. 301 347–349 (2003).
78. Petit, T. *et al.* Probing ion-specific effects on aqueous acetate solutions: Ion pairing versus water structure modifications. *Struct. Dyn.* **1**, (2014).
79. Xie, W., Liu, C., Yang, L. & Gao, Y. On the molecular mechanism of ion specific Hofmeister series. *Sci. China Chem.* **57**, 36–47 (2014).
80. Chen, Y. *et al.* Electrolytes induce long-range orientational order and free energy changes in the H-bond network of bulk water. *Sci. Adv.* **2**, (2016).
81. Liu, C., Min, F., Liu, L. & Chen, J. Hydration properties of alkali and alkaline earth metal ions in aqueous solution: A molecular dynamics study. *Chem. Phys. Lett.* **727**, 31–37 (2019).
82. Smith, J. D., Saykally, R. J. & Geissler, P. L. The effects of dissolved halide anions on hydrogen bonding in liquid water. *J. Am. Chem. Soc.* **129**, 13847–13856 (2007).
83. Stirnemann, G., Wernersson, E., Jungwirth, P. & Laage, D. Mechanisms of Acceleration and Retardation of Water Dynamics by Ions. *J. Am. Chem. Soc.* **135**, 11824–11831 (2013).
84. Laptinskiy, K. A. *et al.* The energy of hydrogen bonds in aqueous suspensions of nanodiamonds with different surface functionalization. *J. Raman Spectrosc.* **50**, 387–395 (2019).
85. Ahmed, M., Namboodiri, V., Singh, A. K. & Mondal, J. A. On the intermolecular vibrational coupling, hydrogen bonding, and librational freedom of water in the hydration shell of mono- and bivalent anions. *J. Chem. Phys.* **141**, (2014).
86. Ahmed, M., Singh, A. K. & Mondal, J. A. Hydrogen-bonding and vibrational coupling of water in a hydrophobic hydration shell as observed by Raman-MCR and isotopic dilution spectroscopy. *Phys. Chem. Chem. Phys.* **18**, 2767–2775 (2016).

87. Ahmed, M., Namboodiri, V., Singh, A. K., Mondal, J. A. & Sarkar, S. K. How ions affect the structure of water: A combined raman spectroscopy and multivariate curve resolution study. *J. Phys. Chem. B* **117**, 16479–16485 (2013).
88. Li, F. & Skinner, J. L. Infrared and Raman line shapes for ice Ih. I. Dilute HOD in H2O and D2O. *J. Chem. Phys.* **132**, (2010).
89. Tomobe, K. *et al.* Origin of the blueshift of water molecules at interfaces of hydrophilic cyclic compounds. *Sci. Adv.* **3**, (2017).
90. Davis, J. G., Gierszal, K. P., Wang, P. & Ben-Amotz, D. Water structural transformation at molecular hydrophobic interfaces. *Nature* **491**, 582–585 (2012).
91. Choe, C., Schleusener, J., Lademann, J. & Darvin, M. E. Keratin-water-NMF interaction as a three layer model in the human stratum corneum using in vivo confocal Raman microscopy. *Sci. Rep.* **7**, (2017).
92. Saberi Movahed, F., Cheng, G. C. & Venkatachari, B. S. Atomistic simulation of thermal decomposition of crosslinked and non-crosslinked phenolic resin chains. in *42nd AIAA Thermophysics Conference* (2011).
93. Brenner, D. *et al.* Nanodiamond-based Nanolubricants: Experiment and Modeling. in *Mater. Res. Soc. Symp. Proc* 1703 (2014). doi:10.1557/opl.2014.
94. Rapaport, D. C. Hydrogen bonds in water Network organization and lifetimes. *Mol. Phys.* **50**, 1151–1162 (1983).
95. Chowdhuri, S. & Chandra, A. Dynamics of halide ion-water hydrogen bonds in aqueous solutions: Dependence on ion size and temperature. *J. Phys. Chem. B* **110**, 9674–9680 (2006).
96. Laage, D., Stirnemann, G., Sterpone, F., Rey, R. & Hynes, J. T. Reorientation and Allied Dynamics in Water and Aqueous Solutions. *Annu. Rev. Phys. Chem.* **62**, 395–416 (2011).
97. Laage, D. & Hynes, J. T. Reorientional dynamics of water molecules in anionic hydration shells. *PNAS* **104**, 11167–11172 (2007).
98. Laage, D. & Hynes, J. T. On the Molecular Mechanism of Water Reorientation. *J. Phys. Chem. B* **112**, 14230–14242 (2008).
99. Tan, H. S., Piletic, I. R. & Fayer, M. D. Orientational dynamics of water confined on a nanometer length scale in reverse micelles. *J. Chem. Phys.* **122**, (2005).
100. Stirnemann, G., Hynes, J. T. & Laage, D. Water Hydrogen Bond Dynamics in Aqueous Solutions of Amphiphiles. *J. Phys. Chem. B* **114**, 3052–3059 (2010).
101. Su, L., Krim, J. & Brenner, D. W. Interdependent Roles of Electrostatics and Surface Functionalization on the Adhesion Strengths of Nanodiamonds to Gold in Aqueous Environments Revealed by Molecular Dynamics Simulations. *J. Phys. Chem. Lett.* **9**, 4396–4400 (2018).
102. Michaud-Agrawal, N., Denning, E. J., Woolf, T. B. & Beckstein, O. MDAnalysis: A toolkit for the analysis of molecular dynamics simulations. *J. Comput. Chem.* **32**, 2319–2327 (2011).
103. Gowers, R. J. *et al.* MDAnalysis: A Python Package for the Rapid Analysis of Molecular Dynamics Simulations. in *Proc. Of The 15th Python In Science Conf. (SciPy 2016)* (2016).
104. Kumar, R., Schmidt, J. R. & Skinner, J. L. Hydrogen bonding definitions and dynamics in liquid water. *J. Chem. Phys.* **126**, (2007).
105. Chowdhuri, S. & Chandra, A. Hydrogen bonds in aqueous electrolyte solutions: Statistics and dynamics based on both geometric and energetic criteria. *Phys. Rev. E* **66**, 7 (2002).
106. Luzar, A. & Chandler, D. Effect of Environment on Hydrogen Bond Dynamics in Liquid

Water Alenka. *Phys. Rev. Lett.* **76**, (1996).
107. Luzar, A. & Chandler, D. Hydrogen-bond kinetics in liquid water. *Nature* **379**, (1996).
108. Kaiser, A., Ritter, M., Nazmutdinov, R. & Probst, M. Hydrogen Bonding and Dielectric Spectra of Ethylene Glycol−Water Mixtures from Molecular Dynamics Simulations. *J. Phys.Chem.B* **120**, 10515–10523 (2016).
109. Kaiser, A. *et al.* Ethylene glycol revisited: Molecular dynamics simulations and visualization of the liquid and its hydrogen-bond network. *J. Mol. Liq.* **189**, 20–29 (2014).
110. Li, Y., Yang, Z., Hu, N., Zhou, R. & Chen, X. Insights into hydrogen bond dynamics at the interface of the charged monolayer-protected Au nanoparticle from molecular dynamics simulation. *J. Chem. Phys* **138**, 184703 (2013).
111. Galamba, N. Water Tetrahedrons, Hydrogen-Bond Dynamics, and the Orientational Mobility of Water around Hydrophobic Solutes. *J. Phys. Chem. B* **118**, 4169–4176 (2014).
112. Raschke, T. M. & Levitt, M. Nonpolar solutes enhance water structure within hydration shells while reducing interactions between them. *PNAS* **10**, 6777–6782 (2005).
113. Sharp, K. A., Madan, B., Manas, E. & Vanderkooi, J. M. Water structure changes induced by hydrophobic and polar solutes revealed by simulations and infrared spectroscopy. *J. Chem. Phys.* **114**, 1791–1796 (2001).
114. Chandra, A. Effects of Ion Atmosphere on Hydrogen-Bond Dynamics in Aqueous Electrolyte Solutions. *Phys. Rev. Lett.* **85**, (2000).
115. Tielrooij, K. J., Garcia-Araez, N., Bonn, M. & Bakker, H. J. Cooperativity in Ion Hydration. *Science* vol. 328 1006–1009 (2010).
116. Tielrooij, K. J., Van Der Post, S. T., Hunger, J., Bonn, M. & Bakker, H. J. Anisotropic water reorientation around ions. *J. Phys. Chem. B* **115**, 12638–12647 (2011).
117. Post, S. T. V. Der & Bakker, H. J. The combined effect of cations and anions on the dynamics of water. *Phys. Chem. Chem. Phys.* **14**, 6280–6288 (2012).
118. Pastorczak, M., Van Der Post, S. T. & Bakker, H. J. Cooperative hydration of carboxylate groups with alkali cations. *Phys.Chem.Chem.Phys* **15**, 17767–17770 (2013).
119. Saberi Movahed, F. & Brenner, D. W. Impacts of surface chemistry and adsorbed ions on dynamics of water around detonation nanodiamond in aqueous salt solutions. *arXiv:2102.13312 [physics.comp-ph]* (2021).
120. Mermoux, M., Crisci, A., Petit, T., Girard, H. A. & Arnault, J.-C. Surface Modifications of Detonation Nanodiamonds Probed by Multiwavelength Raman Spectroscopy. *J.Phys.Chem.C* **118**, 23415–23425 (2014).

# Supplementary Information

In this section, we have presented the tabulated values of parameters associated with HBs of water, which were discussed in the main part of the paper. The mean values of these parameters as well as their 95% confidence interval (CI) are given. The aforementioned statistics have been obtained from five independent MD trajectories. Due to the limited sample size, we have employed the bootstrap percentile method[1], where the bootstrap distribution of a specific statistic was generated by resampling the initial sample data 10,000 times with replacement.

Throughout this document "bulk" water represents the region outside the whole hydration layer of the neutral DND–H that has been solvated in different aqueous salt solutions. Furthermore, the net charge of DNDs is shown with "q".

## S.1 Statistics of HBs' population

### S.1.1 Whole hydration shell of DNDs

**Table S.1.** Different statistics related to the number of water-water HBs for water in the whole hydration shell of DND–H with various surface chemistries and in different salt solutions. $\bar{N}_{HB}$ Denotes the average number of HBs per each water molecule and $N_{HB}^{(i)}$ represents the proportion of water molecules with "$i$" number of water-water HBs per each molecule. The CIs of these statistics are enclosed in parentheses.

| | Salt | $\bar{N}_{HB}$ | $N_{HB}^{(0)}$ | $N_{HB}^{(1)}$ | $N_{HB}^{(2)}$ | $N_{HB}^{(3)}$ | $N_{HB}^{(4)}$ |
|---|---|---|---|---|---|---|---|
| **q = 0** | KCl | 2.977 (2.9744, 2.9800) | 0.50% (0.50%, 0.50%) | 6.10% (6.04%, 6.16%) | 22.26% (22.20%, 22.34%) | 37.40% (37.34%, 37.46%) | 33.72% (33.62%, 33.82%) |
| | NaCl | 2.9734 (2.9704, 2.9760) | 0.50% (0.50%, 0.50%) | 6.16% (6.08%, 6.20%) | 22.32% (22.22%, 22.46%) | 37.38% (37.34%, 37.40%) | 33.60% (33.44%, 33.74%) |
| | CaCl$_2$ | 2.9724 (2.9650, 2.9788) | 0.52% (0.50%, 0.56%) | 6.14% (6.02%, 6.28%) | 22.34% (22.26%, 22.44%) | 37.34% (37.26%, 37.44%) | 33.60% (33.38%, 33.80%) |
| | MgCl$_2$ | 2.9624 (2.9542, 2.9706) | 0.52% (0.50%, 0.56%) | 6.36% (6.16%, 6.56%) | 22.60% (22.40%, 22.80%) | 37.16% (37.06%, 37.26%) | 33.30% (33.02%, 33.58%) |
| **q = +28** | KCl | 2.9888 (2.9806, 2.9942) | 0.52% (0.50%, 0.56%) | 6.14% (6.06%, 6.24%) | 21.08% (20.96%, 21.24%) | 38.22% (38.20%, 38.26%) | 33.98% (33.66%, 34.20%) |
| | NaCl | 2.9888 (2.9810, 2.9962) | 0.52% (0.50%, 0.56%) | 6.12% (6.02%, 6.22%) | 21.10% (20.90%, 21.32%) | 38.24% (38.16%, 38.34%) | 33.96% (33.64%, 34.22%) |
| | CaCl$_2$ | 2.9694 (2.9640, 2.9742) | 0.58% (0.54%, 0.60%) | 6.32% (6.24%, 6.42%) | 21.66% (21.54%, 21.84%) | 38.22% (38.20%, 38.26%) | 33.16% (32.92%, 33.32%) |
| | MgCl$_2$ | 2.9738 (2.9666, 2.9802) | 0.56% (0.52%, 0.60%) | 6.32% (6.22%, 6.46%) | 21.58% (21.42%, 21.74%) | 38.10% (37.94%, 38.24%) | 33.40% (33.22%, 33.58%) |
| **q = +56** | KCl | 2.8458 (2.8384, 2.8532) | 0.70% (0.70%, 0.70%) | 7.30% (7.22%, 7.38%) | 24.90% (24.68%, 25.10%) | 40.68% (40.58%, 40.76%) | 26.36% (26.10%, 26.64%) |
| | NaCl | 2.855 (2.8476, 2.8632) | 0.70% (0.70%, 0.70%) | 7.18% (7.08%, 7.28%) | 24.70% (24.44%, 24.96%) | 40.68% (40.54%, 40.78%) | 26.72% (26.46%, 27.00%) |
| | CaCl$_2$ | 2.825 (2.8164, 2.8332) | 0.70% (0.64%, 0.76%) | 7.46% (7.30%, 7.62%) | 25.82% (25.60%, 26.08%) | 40.36% (40.14%, 40.54%) | 25.58% (25.36%, 25.80%) |
| | MgCl$_2$ | 2.8266 (2.8200, 2.8360) | 0.70% (0.70%, 0.70%) | 7.52% (7.32%, 7.74%) | 25.70% (25.38%, 25.90%) | 40.26% (40.10%, 40.40%) | 25.74% (25.54%, 26.08%) |

| | Salt | $\bar{N}_{HB}$ | $N_{HB}^{(0)}$ | $N_{HB}^{(1)}$ | $N_{HB}^{(2)}$ | $N_{HB}^{(3)}$ | $N_{HB}^{(4)}$ |
|---|---|---|---|---|---|---|---|
| q = +84 | KCl | 2.6732 (2.6646, 2.6808) | 1.06% (1.02%, 1.10%) | 10.08% (9.94%, 10.22%) | 30.00% (29.70%, 30.32%) | 37.80% (37.64%, 37.94%) | 20.96% (20.66%, 21.22%) |
| | NaCl | 2.6708 (2.6614, 2.6802) | 1.10% (1.04%, 1.16%) | 10.10% (9.92%, 10.28%) | 30.20% (30.02%, 30.38%) | 37.74% (37.58%, 37.90%) | 20.84% (20.60%, 21.08%) |
| | CaCl$_2$ | 2.6518 (2.6432, 2.6618) | 1.10% (1.04%, 1.16%) | 10.42% (10.18%, 10.64%) | 30.80% (30.60%, 31.02%) | 37.32% (37.14%, 37.48%) | 20.30% (20.04%, 20.60%) |
| | MgCl$_2$ | 2.6472 (2.6428, 2.6528) | 1.12% (1.10%, 1.16%) | 10.50% (10.38%, 10.62%) | 30.94% (30.82%, 31.00%) | 37.26% (37.22%, 37.30%) | 20.14% (20.00%, 20.34%) |

**Table S.2.** Same as Table S.1, but for the case of DND–NH$_2$.

| | Salt | $\bar{N}_{HB}$ | $N_{HB}^{(0)}$ | $N_{HB}^{(1)}$ | $N_{HB}^{(2)}$ | $N_{HB}^{(3)}$ | $N_{HB}^{(4)}$ |
|---|---|---|---|---|---|---|---|
| q = 0 | KCl | 2.8516 (2.8492, 2.8534) | 0.54% (0.50%, 0.58%) | 6.92% (6.86%, 6.98%) | 25.88% (25.78%, 25.98%) | 40.00% (39.92%, 40.08%) | 26.62% (26.56%, 26.68%) |
| | NaCl | 2.8498 (2.8434, 2.8566) | 0.58% (0.54%, 0.60%) | 6.96% (6.82%, 7.08%) | 25.88% (25.68%, 26.08%) | 39.90% (39.82%, 39.98%) | 26.64% (26.40%, 26.88%) |
| | CaCl$_2$ | 2.841 (2.8364, 2.8470) | 0.60% (0.60%, 0.60%) | 7.20% (7.08%, 7.32%) | 26.04% (25.86%, 26.20%) | 39.66% (39.52%, 39.82%) | 26.46% (26.32%, 26.64%) |
| | MgCl$_2$ | 2.8386 (2.8326, 2.8430) | 0.62% (0.60%, 0.66%) | 7.18% (7.10%, 7.30%) | 26.14% (26.10%, 26.18%) | 39.68% (39.58%, 39.76%) | 26.34% (26.22%, 26.44%) |
| q = +28 | KCl | 2.8108 (2.8018, 2.8198) | 0.70% (0.64%, 0.76%) | 7.82% (7.68%, 7.98%) | 26.78% (26.56%, 27.02%) | 38.78% (38.68%, 38.90%) | 25.84% (25.54%, 26.12%) |
| | NaCl | 2.8156 (2.8110, 2.8206) | 0.68% (0.64%, 0.70%) | 7.72% (7.60%, 7.82%) | 26.80% (26.66%, 26.94%) | 38.88% (38.80%, 38.96%) | 25.90% (25.76%, 26.08%) |
| | CaCl$_2$ | 2.791 (2.7810, 2.8008) | 0.76% (0.72%, 0.80%) | 8.16% (7.94%, 8.38%) | 27.38% (27.22%, 27.60%) | 38.54% (38.24%, 38.78%) | 25.14% (24.80%, 25.42%) |
| | MgCl$_2$ | 2.789 (2.7830, 2.7966) | 0.76% (0.72%, 0.80%) | 8.24% (8.10%, 8.34%) | 27.38% (27.28%, 27.50%) | 38.42% (38.36%, 38.48%) | 25.16% (24.94%, 25.42%) |
| q = +56 | KCl | 2.7052 (2.7000, 2.7108) | 1.00% (0.94%, 1.06%) | 10.02% (9.88%, 10.14%) | 29.62% (29.54%, 29.72%) | 36.02% (35.86%, 36.14%) | 23.30% (23.10%, 23.50%) |
| | NaCl | 2.71 (2.7020, 2.7158) | 0.98% (0.92%, 1.04%) | 9.88% (9.76%, 10.04%) | 29.62% (29.48%, 29.78%) | 36.04% (35.86%, 36.22%) | 23.44% (23.24%, 23.64%) |
| | CaCl$_2$ | 2.694 (2.6872, 2.7032) | 1.02% (1.00%, 1.06%) | 10.10% (9.88%, 10.30%) | 29.94% (29.86%, 30.04%) | 36.02% (35.86%, 36.20%) | 22.84% (22.64%, 23.04%) |
| | MgCl$_2$ | 2.6864 (2.6722, 2.6986) | 1.10% (1.04%, 1.16%) | 10.34% (10.02%, 10.66%) | 30.04% (29.78%, 30.30%) | 35.78% (35.62%, 35.92%) | 22.72% (22.28%, 23.04%) |
| q = +84 | KCl | 2.593 (2.5832, 2.6016) | 1.56% (1.46%, 1.66%) | 12.62% (12.42%, 12.88%) | 31.04% (30.90%, 31.20%) | 34.20% (34.10%, 34.28%) | 20.50% (20.20%, 20.76%) |
| | NaCl | 2.5962 (2.5838, 2.6074) | 1.64% (1.52%, 1.76%) | 12.68% (12.44%, 12.92%) | 30.62% (30.46%, 30.86%) | 34.38% (34.22%, 34.56%) | 20.64% (20.34%, 20.92%) |
| | CaCl$_2$ | 2.571 (2.5642, 2.5772) | 1.66% (1.62%, 1.70%) | 13.06% (12.86%, 13.22%) | 31.44% (31.32%, 31.56%) | 33.96% (33.76%, 34.12%) | 19.82% (19.68%, 19.98%) |
| | MgCl$_2$ | 2.5758 (2.5680, 2.5826) | 1.64% (1.60%, 1.68%) | 13.06% (12.82%, 13.30%) | 31.18% (31.12%, 31.24%) | 34.00% (33.94%, 34.06%) | 20.04% (19.80%, 20.26%) |

**Table S.3.** Same as Table S.1, but for the case of DND–COOH.

| | Salt | $\bar{N}_{HB}$ | $N_{HB}^{(0)}$ | $N_{HB}^{(1)}$ | $N_{HB}^{(2)}$ | $N_{HB}^{(3)}$ | $N_{HB}^{(4)}$ |
|---|---|---|---|---|---|---|---|
| **q = 0** | KCl | 2.6368 (2.6316, 2.6430) | 0.82% (0.80%, 0.86%) | 10.02% (9.90%, 10.10%) | 31.80% (31.60%, 31.96%) | 39.14% (39.02%, 39.28%) | 18.16% (18.02%, 18.34%) |
| | NaCl | 2.6504 (2.6456, 2.6560) | 0.84% (0.80%, 0.88%) | 9.78% (9.68%, 9.88%) | 31.34% (31.22%, 31.40%) | 39.50% (39.38%, 39.66%) | 18.52% (18.40%, 18.62%) |
| | CaCl$_2$ | 2.6366 (2.6250, 2.6458) | 0.94% (0.86%, 1.04%) | 10.18% (9.96%, 10.44%) | 31.36% (31.22%, 31.48%) | 39.08% (38.90%, 39.26%) | 18.38% (18.14%, 18.58%) |
| | MgCl$_2$ | 2.655 (2.6470, 2.6638) | 0.78% (0.74%, 0.80%) | 9.68% (9.50%, 9.84%) | 31.32% (31.16%, 31.48%) | 39.62% (39.46%, 39.82%) | 18.58% (18.38%, 18.78%) |
| **q = −28** | KCl | 2.6058 (2.6012, 2.6102) | 1.04% (1.00%, 1.08%) | 10.86% (10.80%, 10.94%) | 32.32% (32.16%, 32.44%) | 37.88% (37.78%, 37.96%) | 17.86% (17.74%, 17.98%) |
| | NaCl | 2.6186 (2.6122, 2.6244) | 1.08% (1.02%, 1.14%) | 10.76% (10.56%, 10.96%) | 31.78% (31.70%, 31.86%) | 37.74% (37.54%, 37.94%) | 18.58% (18.48%, 18.70%) |
| | CaCl$_2$ | 2.5878 (2.5802, 2.5954) | 1.36% (1.26%, 1.44%) | 11.86% (11.72%, 11.98%) | 31.62% (31.46%, 31.76%) | 36.72% (36.58%, 36.84%) | 18.38% (18.18%, 18.58%) |
| | MgCl$_2$ | 2.5664 (2.5610, 2.5706) | 1.48% (1.38%, 1.56%) | 12.00% (11.84%, 12.16%) | 32.34% (32.20%, 32.48%) | 36.52% (36.42%, 36.62%) | 17.60% (17.42%, 17.74%) |
| **q = −56** | KCl | 2.5374 (2.5312, 2.5424) | 1.38% (1.32%, 1.44%) | 12.68% (12.56%, 12.82%) | 33.42% (33.26%, 33.54%) | 35.78% (35.62%, 35.92%) | 16.72% (16.60%, 16.80%) |
| | NaCl | 2.5842 (2.5776, 2.5908) | 1.30% (1.22%, 1.38%) | 11.82% (11.68%, 11.98%) | 32.42% (32.40%, 32.46%) | 36.08% (35.92%, 36.22%) | 18.38% (18.26%, 18.50%) |
| | CaCl$_2$ | 2.4964 (2.4874, 2.5038) | 2.24% (2.12%, 2.36%) | 14.08% (13.82%, 14.42%) | 32.60% (32.52%, 32.68%) | 33.72% (33.54%, 33.88%) | 17.30% (17.12%, 17.48%) |
| | MgCl$_2$ | 2.4606 (2.4588, 2.4626) | 2.66% (2.56%, 2.74%) | 14.14% (13.88%, 14.38%) | 33.78% (33.64%, 33.94%) | 33.08% (32.98%, 33.20%) | 16.28% (16.18%, 16.36%) |
| **q = −84** | KCl | 2.4938 (2.4862, 2.5014) | 1.80% (1.68%, 1.92%) | 13.72% (13.62%, 13.86%) | 33.62% (33.60%, 33.66%) | 34.78% (34.62%, 34.96%) | 16.02% (15.88%, 16.14%) |
| | NaCl | 2.5216 (2.5144, 2.5280) | 1.82% (1.72%, 1.96%) | 13.20% (13.04%, 13.32%) | 33.14% (33.10%, 33.18%) | 34.44% (34.32%, 34.58%) | 17.34% (17.22%, 17.44%) |
| | CaCl$_2$ | 2.4094 (2.3982, 2.4200) | 3.46% (3.26%, 3.64%) | 15.64% (15.42%, 15.98%) | 33.24% (33.14%, 33.38%) | 31.42% (31.14%, 31.64%) | 16.14% (15.98%, 16.30%) |
| | MgCl$_2$ | 2.4072 (2.4024, 2.4124) | 3.68% (3.62%, 3.74%) | 15.04% (14.96%, 15.14%) | 33.84% (33.66%, 33.98%) | 31.44% (31.32%, 31.50%) | 15.92% (15.78%, 16.04%) |

**Table S.4.** Same as Table S.1, but for the case of DND–OH.

| | Salt | $\bar{N}_{HB}$ | $N_{HB}^{(0)}$ | $N_{HB}^{(1)}$ | $N_{HB}^{(2)}$ | $N_{HB}^{(3)}$ | $N_{HB}^{(4)}$ |
|---|---|---|---|---|---|---|---|
| **q = 0** | KCl | 2.8332 (2.8264, 2.8400) | 0.58% (0.54%, 0.60%) | 7.26% (7.12%, 7.40%) | 26.60% (26.44%, 26.76%) | 39.22% (39.16%, 39.28%) | 26.30% (26.06%, 26.54%) |
| | NaCl | 2.8364 (2.8314, 2.8410) | 0.58% (0.54%, 0.60%) | 7.22% (7.12%, 7.32%) | 26.46% (26.32%, 26.56%) | 39.14% (39.06%, 39.20%) | 26.52% (26.32%, 26.68%) |

|  | 2.8256 | 0.66% | 7.52% | 26.64% | 38.80% | 26.34% |
| --- | --- | --- | --- | --- | --- | --- |
| CaCl$_2$ | (2.8178, 2.8364) | (0.62%, 0.70%) | (7.24%, 7.72%) | (26.46%, 26.80%) | (38.60%, 39.08%) | (26.16%, 26.58%) |
| MgCl$_2$ | 2.8252 | 0.60% | 7.48% | 26.84% | 38.88% | 26.18% |
|  | (2.8154, 2.8334) | (0.60%, 0.60%) | (7.34%, 7.62%) | (26.62%, 27.14%) | (38.70%, 39.06%) | (25.88%, 26.40%) |

### S.1.2 Bulk water

**Table S.5.** Same as Table S.1, but for the case of bulk water.

| Salt | $\bar{N}_{HB}$ | $N_{HB}^{(0)}$ | $N_{HB}^{(1)}$ | $N_{HB}^{(2)}$ | $N_{HB}^{(3)}$ | $N_{HB}^{(4)}$ |
| --- | --- | --- | --- | --- | --- | --- |
| KCl | 3.017 | 0.50% | 5.80% | 21.50% | 35.50% | 36.60% |
|  | (3.0170, 3.0170) | (0.50%, 0.50%) | (5.80%, 5.80%) | (21.50%, 21.50%) | (35.50%, 35.50%) | (36.60%, 36.60%) |
| NaCl | 3.016 | 0.50% | 5.90% | 21.60% | 35.50% | 36.50% |
|  | (3.0160, 3.0160) | (0.50%, 0.50%) | (5.90%, 5.90%) | (21.60%, 21.60%) | (35.50%, 35.50%) | (36.50%, 36.50%) |
| CaCl$_2$ | 3.002 | 0.50% | 6.10% | 21.90% | 35.30% | 36.10% |
|  | (3.0020, 3.0020) | (0.50%, 0.50%) | (6.10%, 6.10%) | (21.90%, 21.90%) | (35.30%, 35.30%) | (36.10%, 36.10%) |
| MgCl$_2$ | 3.005 | 0.50% | 6.00% | 21.90% | 35.30% | 36.20% |
|  | (3.0050, 3.0050) | (0.50%, 0.50%) | (6.00%, 6.00%) | (21.90%, 21.90%) | (35.30%, 35.30%) | (36.20%, 36.20%) |

### S.1.3 Whole hydration shell of ions

**Table S.6.** Same as Table S.1, but for the case of ions that are dissolved in the aqueous solution of neutral DND–H.

| Ion | $\bar{N}_{HB}$ | $N_{HB}^{(0)}$ | $N_{HB}^{(1)}$ | $N_{HB}^{(2)}$ | $N_{HB}^{(3)}$ | $N_{HB}^{(4)}$ |
| --- | --- | --- | --- | --- | --- | --- |
| K$^+$ | 2.0446 | 3.02% | 23.32% | 42.96% | 27.58% | 3.12% |
|  | (2.0362, 2.0522) | (2.94%, 3.12%) | (23.06%, 23.62%) | (42.82%, 43.06%) | (27.30%, 27.82%) | (3.10%, 3.16%) |
| Na$^+$ | 1.918 | 3.48% | 26.64% | 46.16% | 22.12% | 1.62% |
|  | (1.9126, 1.9244) | (3.44%, 3.50%) | (26.46%, 26.78%) | (46.06%, 46.26%) | (21.98%, 22.28%) | (1.56%, 1.68%) |
| Ca$^{2+}$ | 1.7232 | 3.48% | 32.20% | 53.48% | 10.28% | 0.58% |
|  | (1.7178, 1.7282) | (3.42%, 3.54%) | (31.98%, 32.42%) | (53.28%, 53.68%) | (10.18%, 10.38%) | (0.54%, 0.60%) |
| Mg$^{2+}$ | 1.529 | 2.96% | 44.62% | 49.14% | 3.20% | 0.10% |
|  | (1.5220, 1.5350) | (2.86%, 3.08%) | (44.34%, 44.90%) | (48.74%, 49.52%) | (3.14%, 3.26%) | (0.10%, 0.10%) |
| Cl$^-$ (NaCl) | 2.1018 | 2.38% | 21.88% | 42.40% | 29.78% | 3.54% |
|  | (2.0948, 2.1088) | (2.32%, 2.44%) | (21.70%, 22.06%) | (42.34%, 42.46%) | (29.58%, 29.98%) | (3.46%, 3.64%) |
| Cl$^-$ (MgCl$_2$) | 2.137 | 2.40% | 21.12% | 41.24% | 30.70% | 4.50% |
|  | (2.1320, 2.1426) | (2.34%, 2.46%) | (20.98%, 21.28%) | (41.20%, 41.28%) | (30.54%, 30.86%) | (4.44%, 4.56%) |

## S.2 Dynamics of HBs

Throughout this section the characteristic times $\tau^{(c)}$ and $\tau^{(i)}$ for HBs of water with different species are presented. The former and the latter represent the lifetime and the structural relaxation time of HBs.

### S.2.1 Water–Water HBs

#### S.2.1.1 Whole hydration shell of DNDs

**Table S.7.** The average values of $\tau^{(c)}$ and $\tau^{(i)}$ of water-water HBs for water in the whole hydration layer of DND–H with various surface chemistries and in different salt solutions. The CIs of the mean values are enclosed in parentheses.

| | Salt | $\tau^{(c)}$ / ps | $\tau^{(i)}$ / ps | | Salt | $\tau^{(c)}$ / ps | $\tau^{(i)}$ / ps |
|---|---|---|---|---|---|---|---|
| q = 0 | KCl | 1.179 (1.1756, 1.1825) | 4.1941 (4.1758, 4.2124) | q = +28 | KCl | 1.5276 (1.5060, 1.5454) | 5.4951 (5.3728, 5.6113) |
| | NaCl | 1.1735 (1.1674, 1.1800) | 4.1842 (4.1461, 4.2284) | | NaCl | 1.5427 (1.5358, 1.5478) | 5.5554 (5.5312, 5.5723) |
| | CaCl$_2$ | 1.1823 (1.1725, 1.1917) | 4.2084 (4.1556, 4.2622) | | CaCl$_2$ | 1.5389 (1.5318, 1.5456) | 5.5438 (5.4788, 5.6088) |
| | MgCl$_2$ | 1.187 (1.1841, 1.1910) | 4.2475 (4.2216, 4.2800) | | MgCl$_2$ | 1.5474 (1.5227, 1.5704) | 5.5661 (5.4388, 5.6665) |
| q = +56 | KCl | 1.6237 (1.6132, 1.6345) | 5.4113 (5.3374, 5.4790) | q = +84 | KCl | 1.3716 (1.3633, 1.3791) | 4.7576 (4.7014, 4.8139) |
| | NaCl | 1.6054 (1.5949, 1.6171) | 5.3529 (5.2917, 5.4141) | | NaCl | 1.3637 (1.3537, 1.3755) | 4.7558 (4.7118, 4.7998) |
| | CaCl$_2$ | 1.6268 (1.6142, 1.6393) | 5.4323 (5.3298, 5.5257) | | CaCl$_2$ | 1.3637 (1.3555, 1.3720) | 4.7101 (4.6457, 4.7948) |
| | MgCl$_2$ | 1.6355 (1.6175, 1.6577) | 5.5089 (5.4036, 5.6393) | | MgCl$_2$ | 1.3757 (1.3636, 1.3877) | 4.7879 (4.7182, 4.8576) |

**Table S.8.** Same as Table S.7, but for the case of DND–NH$_2$.

| | Salt | $\tau^{(c)}$ / ps | $\tau^{(i)}$ / ps | | Salt | $\tau^{(c)}$ / ps | $\tau^{(i)}$ / ps |
|---|---|---|---|---|---|---|---|
| q = 0 | KCl | 1.4894 (1.4779, 1.5030) | 5.5469 (5.4679, 5.6502) | q = +28 | KCl | 1.4719 (1.4527, 1.4884) | 5.5684 (5.4767, 5.6459) |
| | NaCl | 1.5007 (1.4885, 1.5133) | 5.6845 (5.6234, 5.7572) | | NaCl | 1.4587 (1.4473, 1.4697) | 5.478 (5.4071, 5.5452) |
| | CaCl$_2$ | 1.496 (1.4849, 1.5091) | 5.6576 (5.6093, 5.7169) | | CaCl$_2$ | 1.4568 (1.4514, 1.4618) | 5.481 (5.4508, 5.5166) |
| | MgCl$_2$ | 1.5206 (1.5123, 1.5294) | 5.7582 (5.6861, 5.8392) | | MgCl$_2$ | 1.4657 (1.4516, 1.4821) | 5.5358 (5.4665, 5.6303) |
| q = +56 | KCl | 1.4106 (1.4012, 1.4225) | 5.3969 (5.3352, 5.4722) | q = +84 | KCl | 1.4545 (1.4425, 1.4671) | 6.1344 (6.0784, 6.1975) |
| | NaCl | 1.405 (1.4002, 1.4100) | 5.3751 (5.3328, 5.4110) | | NaCl | 1.4551 (1.4519, 1.4602) | 6.0668 (6.0254, 6.0987) |
| | CaCl$_2$ | 1.4102 (1.4026, 1.4228) | 5.3974 (5.3328, 5.4640) | | CaCl$_2$ | 1.4329 (1.4219, 1.4463) | 5.9192 (5.8334, 5.9992) |
| | MgCl$_2$ | 1.4067 (1.4053, 1.4082) | 5.4024 (5.3662, 5.4359) | | MgCl$_2$ | 1.4265 (1.4136, 1.4370) | 5.9349 (5.8652, 6.0221) |

**Table S.9.** Same as Table S.7, but for the case of DND–COOH.

| | Salt | $\tau^{(c)}$ / ps | $\tau^{(i)}$ / ps | | Salt | $\tau^{(c)}$ / ps | $\tau^{(i)}$ / ps |
|---|---|---|---|---|---|---|---|
| **q = 0** | KCl | 1.1702 (1.1652, 1.1750) | 4.6293 (4.5931, 4.6751) | **q = −28** | KCl | 1.1718 (1.1603, 1.1834) | 4.5959 (4.5371, 4.6547) |
| | NaCl | 1.1643 (1.1538, 1.1743) | 4.5747 (4.5106, 4.6387) | | NaCl | 1.177 (1.1746, 1.1792) | 4.6133 (4.5717, 4.6519) |
| | CaCl$_2$ | 1.1892 (1.1834, 1.1966) | 4.7352 (4.6708, 4.7996) | | CaCl$_2$ | 1.2061 (1.1951, 1.2159) | 4.8574 (4.7416, 4.9596) |
| | MgCl$_2$ | 1.1835 (1.1751, 1.1903) | 4.699 (4.6407, 4.7436) | | MgCl$_2$ | 1.3031 (1.2877, 1.3211) | 5.5023 (5.4213, 5.6017) |
| **q = −56** | KCl | 1.1567 (1.1504, 1.1621) | 4.7287 (4.6757, 4.7835) | **q = −84** | KCl | 1.1748 (1.1661, 1.1840) | 5.0137 (4.9867, 5.0408) |
| | NaCl | 1.1531 (1.1498, 1.1556) | 4.7688 (4.7419, 4.7957) | | NaCl | 1.1695 (1.1653, 1.1739) | 5.1167 (5.0309, 5.1965) |
| | CaCl$_2$ | 1.2359 (1.2304, 1.2441) | 5.4636 (5.4044, 5.5362) | | CaCl$_2$ | 1.2976 (1.2880, 1.3065) | 6.383 (6.2551, 6.5024) |
| | MgCl$_2$ | 1.408 (1.3925, 1.4233) | 6.5311 (6.4284, 6.6272) | | MgCl$_2$ | 1.4894 (1.4692, 1.5082) | 7.5095 (7.3017, 7.6813) |

**Table S.10.** Same as Table S.7, but for the case of DND–OH.

| | Salt | $\tau^{(c)}$ / ps | $\tau^{(i)}$ / ps |
|---|---|---|---|
| **q = 0** | KCl | 1.3623 (1.3552, 1.3693) | 4.9277 (4.8891, 4.9662) |
| | NaCl | 1.3653 (1.3512, 1.3794) | 4.9406 (4.8693, 5.0228) |
| | CaCl$_2$ | 1.3809 (1.3754, 1.3854) | 5.0364 (4.9881, 5.0753) |
| | MgCl$_2$ | 1.3819 (1.3756, 1.3875) | 5.0263 (4.9591, 5.0842) |

### S.2.1.2 Bulk water

**Table S.11.** Same as Table S.7, but for the case of bulk water.

| Salt | $\tau^{(c)}$ / ps | $\tau^{(i)}$ / ps |
|---|---|---|
| KCl | 1.1757 (1.1757, 1.1757) | 3.4587 (3.4587, 3.4587) |
| NaCl | 1.1735 (1.1735, 1.1735) | 3.4516 (3.4516, 3.4516) |
| CaCl$_2$ | 1.1893 (1.1893, 1.1893) | 3.5355 (3.5355, 3.5355) |

|  | 1.1948 | 3.5554 |
|---|---|---|
| MgCl$_2$ | (1.1948, 1.1948) | (3.5554, 3.5554) |

### S.2.1.3 Whole hydration shell of ions

**Table S.12.** Same as Table S.7, but for water in the whole hydration shell of ions dissolved in different aqueous solutions of the neutral DND–H.

| Ion | $\tau^{(c)}$ / ps | $\tau^{(i)}$ / ps |
|---|---|---|
| K$^+$ | 0.7825 (0.7778, 0.7864) | 2.2724 (2.2562, 2.2841) |
| Na$^+$ | 0.786 (0.7843, 0.7879) | 2.5256 (2.4966, 2.5443) |
| Ca$^{2+}$ | 1.125 (1.1170, 1.1317) | 3.8266 (3.7982, 3.8552) |
| Mg$^{2+}$ | 2.0152 (2.0011, 2.0303) | 7.6806 (7.5842, 7.7960) |
| Cl$^-$ (NaCl) | 0.9465 (0.9428, 0.9505) | 2.8155 (2.8033, 2.8284) |
| Cl$^-$ (MgCl$_2$) | 0.9819 (0.9793, 0.9858) | 2.9685 (2.9550, 2.9828) |

### S.2.2 Water–Site HBs

Water–Site HBs refer to HBs that are formed between water and the surface functional groups of DNDs.

**Table S.13.** The average values of $\tau^{(c)}$ and $\tau^{(i)}$ of water-site HBs for water at the interface with DND–COOH with various surface chemistries and in different salt solutions. The CIs of the mean values are enclosed in parentheses. "D" and "A" represent, respectively, the donor and acceptor heavy atoms of the hydrogen bonded species. OW, O, OH, and O2 denote, respectively, the oxygen of water, the oxygen in –COO$^-$ functional group, the oxygen in the hydroxyl portion of –COOH functional group, and the oxygen doubly bonded to carbon in –COOH functional group. As a reminder, –COO$^-$ only exists on the surface of charged DND–COOH. Thus, the rows for "D: OW, A: O" water-site HBs corresponding to q = 0 are empty.

|  | Salt | D: OW, A: O | | D: OH, A: OW | | D: OW, A: O2 | |
|---|---|---|---|---|---|---|---|
|  |  | $\tau^{(c)}$ / ps | $\tau^{(i)}$ / ps | $\tau^{(c)}$ / ps | $\tau^{(i)}$ / ps | $\tau^{(c)}$ / ps | $\tau^{(i)}$ / ps |
| q = 0 | KCl | – | – | 6.2554 (6.1568, 6.4003) | 60.9587 (59.3187, 63.1638) | 0.6566 (0.6397, 0.6734) | 7.6083 (6.9373, 8.1339) |
|  | NaCl | – | – | 5.6904 (5.5607, 5.8058) | 55.2355 (53.4002, 58.1633) | 0.6525 (0.6358, 0.6693) | 7.768 (7.5640, 8.0558) |
|  | CaCl$_2$ | – | – | 5.867 (5.7577, 5.9673) | 57.0286 (54.0367, 60.4559) | 0.6695 (0.6602, 0.6792) | 8.2936 (8.0279, 8.6114) |
|  | MgCl$_2$ | – | – | 5.748 (5.5973, 5.8922) | 55.9675 (53.9342, 57.9166) | 0.6493 (0.6262, 0.6780) | 7.9572 (7.7794, 8.1560) |
|  | KCl | 4.1481 (2.7592, 4.9736) | 52.4732 (40.8760, 69.0682) | 4.6627 (4.5344, 4.8073) | 63.9869 (60.0948, 68.6317) | 0.6536 (0.6354, 0.6711) | 7.6592 (7.4564, 7.8791) |

| | Salt | | | | | | |
|---|---|---|---|---|---|---|---|
| q = −28 | NaCl | 3.9684 (3.7725, 4.1867) | 38.0356 (34.9225, 41.3216) | 4.5322 (4.4440, 4.6276) | 71.9804 (66.0041, 79.2162) | 0.6316 (0.6192, 0.6443) | 7.6087 (7.4290, 7.7928) |
| | CaCl$_2$ | 6.0555 (4.5873, 7.1607) | 52.3046 (42.0766, 62.5327) | 6.8945 (6.8059, 6.9822) | 61.4525 (59.9560, 63.3446) | 0.6844 (0.6701, 0.6960) | 8.3544 (7.5221, 9.0308) |
| | MgCl$_2$ | 6.6595 (4.4462, 8.2693) | 79.4422 (67.0850, 91.7664) | 4.5202 (4.4734, 4.5830) | 65.2776 (63.8613, 67.3591) | 0.6787 (0.6666, 0.6908) | 8.6872 (8.2819, 9.0116) |
| q = −56 | KCl | 5.6125 (3.7073, 6.7303) | 65.5674 (55.1356, 77.3841) | 9.1785 (8.8538, 9.6245) | 58.4427 (55.8715, 61.5088) | 0.6475 (0.6336, 0.6614) | 8.9538 (8.3993, 9.5083) |
| | NaCl | 5.3633 (4.8595, 5.7485) | 50.9315 (48.2547, 52.9542) | 8.4042 (8.0836, 8.7231) | 58.2372 (55.2256, 61.4507) | 0.6206 (0.5695, 0.6526) | 9.4464 (9.1491, 9.8214) |
| | CaCl$_2$ | 7.2378 (5.7222, 8.3568) | 77.4147 (69.4963, 89.1954) | 8.3513 (8.1041, 8.6199) | 55.1474 (52.7008, 57.9725) | 0.6706 (0.6573, 0.6832) | 11.0813 (10.6434, 11.4709) |
| | MgCl$_2$ | 9.5284 (7.9229, 10.4630) | 87.87 (69.4632, 103.2553) | 8.6447 (8.4432, 8.8573) | 57.2006 (54.8042, 59.0575) | 0.696 (0.6615, 0.7174) | 11.7207 (11.2088, 12.2326) |
| q = −84 | KCl | 6.5195 (5.3556, 7.2827) | 84.7992 (77.1514, 94.6755) | 9.9685 (9.7708, 10.1693) | 68.5238 (66.8968, 70.7398) | 0.6168 (0.6064, 0.6270) | 8.5679 (8.1469, 9.1672) |
| | NaCl | 6.5764 (6.3042, 6.8416) | 74.9043 (69.0488, 79.8206) | 9.3943 (9.0160, 9.7641) | 62.3884 (61.3758, 63.4010) | 0.6514 (0.6352, 0.6673) | 9.652 (8.8259, 10.5480) |
| | CaCl$_2$ | 9.1539 (8.7066, 9.6013) | 116.9819 (102.9789, 137.0931) | 10.1961 (9.7573, 10.7227) | 64.7911 (59.7072, 69.1977) | 0.7395 (0.7024, 0.7766) | 14.5795 (13.4572, 15.2955) |
| | MgCl$_2$ | 9.6119 (9.0457, 10.2022) | 129.5118 (115.5461, 143.4776) | 8.6328 (8.4138, 8.8580) | 64.7021 (58.4640, 70.9403) | 0.7277 (0.6926, 0.7674) | 13.8588 (12.9300, 14.8444) |

**Table S.14.** Same as Table S.13, but for water at the interface with DND–NH$_2$. Here, N3 and NT denote, respectively, the N atoms in –NH$_3^+$ and –NH$_2$ functional groups. Since –NH$_3^+$ only exist on the surface of the charged DND–NH$_2$, the rows corresponding to "D: N3, A: OW" for the neutral DND are empty.

| | Salt | D: N3, A: OW | | D: OW, A: NT | | D: NT, A: OW | |
|---|---|---|---|---|---|---|---|
| | | $\tau^{(c)}$ / ps | $\tau^{(i)}$ / ps | $\tau^{(c)}$ / ps | $\tau^{(i)}$ / ps | $\tau^{(c)}$ / ps | $\tau^{(i)}$ / ps |
| q = 0 | KCl | – | – | 3.1576 (2.9537, 3.3616) | 22.9346 (22.4786, 23.3905) | 0.6806 (0.6703, 0.6941) | 11.9155 (11.3980, 12.4330) |
| | NaCl | – | – | 3.1535 (3.1220, 3.1850) | 24.9361 (24.1611, 25.7111) | 0.6691 (0.6461, 0.6901) | 12.3371 (11.5255, 13.0838) |
| | CaCl$_2$ | – | – | 3.1009 (3.0968, 3.1050) | 22.2862 (21.6988, 22.8736) | 0.664 (0.6360, 0.6843) | 11.6609 (11.1503, 12.0730) |
| | MgCl$_2$ | – | – | 3.2406 (3.2104, 3.2708) | 23.8578 (21.7801, 25.9356) | 0.6869 (0.6578, 0.7109) | 11.7599 (10.8020, 12.5434) |
| | KCl | 2.1581 (1.8684, 2.4588) | 65.6359 (56.2120, 79.0211) | 2.9005 (2.8329, 2.9680) | 22.7869 (22.6239, 22.9498) | 0.7265 (0.7098, 0.7446) | 11.8357 (11.1768, 12.3888) |
| | NaCl | 2.1182 | 69.2047 | 2.8703 | 20.2854 | 0.7414 | 11.9231 |

|       | Salt  |                   |                     |                  |                    |                  |                     |
|-------|-------|-------------------|---------------------|------------------|--------------------|------------------|---------------------|
| q = +28 |       | (1.7827, 2.4537) | (57.8494, 83.9703) | (2.8502, 2.8904) | (20.0751, 20.4957) | (0.7291, 0.7553) | (11.3743, 12.4791) |
|       | CaCl$_2$ | 1.9257<br>(1.7593, 2.2132) | 51.833<br>(25.9587, 71.4394) | 2.7149<br>(2.6701, 2.7596) | 19.1908<br>(19.1908, 19.1908) | 0.7251<br>(0.7168, 0.7336) | 11.5716<br>(10.5190, 12.4564) |
|       | MgCl$_2$ | 2.3365<br>(2.0661, 2.5486) | 62.5673<br>(54.9457, 74.1639) | 2.7804<br>(2.6771, 2.8838) | 19.7415<br>(19.1166, 20.3664) | 0.7264<br>(0.7171, 0.7326) | 12.1836<br>(11.9338, 12.4333) |
| q = +56 | KCl | 2.7957<br>(2.7171, 2.8753) | 95.82<br>(91.0693, 100.1977) | 1.3755<br>(1.2566, 1.4943) | 9.4693<br>(9.2426, 9.6961) | 0.8112<br>(0.7950, 0.8273) | 14.1781<br>(13.8470, 14.7213) |
|       | NaCl | 2.7082<br>(2.5197, 2.8968) | 95.8875<br>(89.1353, 106.8857) | 1.4083<br>(1.2069, 1.6097) | 9.3313<br>(8.1037, 10.5590) | 0.8029<br>(0.7888, 0.8155) | 13.9824<br>(13.6639, 14.3009) |
|       | CaCl$_2$ | 2.9714<br>(2.8570, 3.0858) | 118.6321<br>(103.0158, 135.5785) | 1.5759<br>(1.5021, 1.6497) | 10.1844<br>(9.3036, 11.0651) | 0.8012<br>(0.7814, 0.8226) | 13.982<br>(13.6621, 14.2827) |
|       | MgCl$_2$ | 2.8685<br>(2.7447, 2.9619) | 103.6073<br>(98.6521, 108.5625) | 1.3864<br>(1.2345, 1.5384) | 10.0991<br>(9.1305, 11.0677) | 0.8209<br>(0.8046, 0.8305) | 14.6591<br>(14.0207, 15.3026) |
| q = +84 | KCl | 4.5155<br>(4.3215, 4.6720) | 129.7083<br>(109.8845, 149.5321) | 1.697<br>(1.6509, 1.7431) | 10.3173<br>(9.5958, 11.0388) | 0.9398<br>(0.9237, 0.9531) | 20.0134<br>(19.0663, 20.9605) |
|       | NaCl | 4.7015<br>(4.5246, 4.8739) | 99.1042<br>(83.0675, 115.9073) | 2.1374<br>(2.0789, 2.1959) | 11.5272<br>(11.3555, 11.6989) | 0.9559<br>(0.9235, 0.9789) | 20.0456<br>(18.7176, 21.3779) |
|       | CaCl$_2$ | 4.392<br>(4.2956, 4.5074) | 95.8375<br>(85.1195, 108.0425) | 1.8861<br>(1.7614, 2.0108) | 11.6317<br>(11.0070, 12.2563) | 0.951<br>(0.9328, 0.9691) | 20.1153<br>(19.4754, 20.9137) |
|       | MgCl$_2$ | 4.3942<br>(4.2380, 4.6046) | 105.177<br>(94.2862, 119.3721) | 1.6938<br>(1.6554, 1.7323) | 11.7574<br>(10.6414, 12.8733) | 0.9389<br>(0.9251, 0.9584) | 20.2107<br>(19.7418, 20.9024) |

**Table S.15.** Same as Table S.13, but for water at the interface with DND–OH. Here, O denote the oxygen atom in the hydroxyl surface functional group.

|       | Salt  | D: OW, A: O | | D: O, A: OW | |
|-------|-------|---|---|---|---|
|       |       | $\tau^{(c)}$ / ps | $\tau^{(i)}$ / ps | $\tau^{(c)}$ / ps | $\tau^{(i)}$ / ps |
| q = 0 | KCl | 1.5118<br>(1.4482, 1.5771) | 11.3479<br>(10.7762, 11.8065) | 1.8255<br>(1.6698, 1.9407) | 18.3784<br>(17.6499, 19.1068) |
|       | NaCl | 1.5566<br>(1.4838, 1.6053) | 11.2382<br>(10.6543, 11.6885) | 1.8345<br>(1.7060, 1.9632) | 18.2279<br>(17.4729, 19.1564) |
|       | CaCl$_2$ | 1.4806<br>(1.4327, 1.5619) | 11.6775<br>(10.9559, 12.2790) | 1.8142<br>(1.7220, 1.8784) | 18.4761<br>(17.6639, 19.0749) |
|       | MgCl$_2$ | 1.5811<br>(1.5099, 1.6406) | 10.9101<br>(10.5423, 11.2258) | 1.7588<br>(1.6171, 1.9006) | 17.9888<br>(17.3088, 18.7929) |

### S.2.3 Water–Anion HBs

The results for HBs that are donated by water molecules to Cl⁻ anion in different aqueous salt solutions of various DNDs are presented.

**Table S.16.** The average values of $\tau^{(c)}$ and $\tau^{(i)}$ of water-anion HBs for water in different aqueous salt solutions of DND–H with various surface chemistries. The CIs of the mean values are enclosed in parentheses.

| | Salt | $\tau^{(c)}$ / ps | $\tau^{(i)}$ / ps | | Salt | $\tau^{(c)}$ / ps | $\tau^{(i)}$ / ps |
|---|---|---|---|---|---|---|---|
| q = 0 | KCl | 2.2625 (2.2543, 2.2707) | 10.3821 (10.3530, 10.4111) | q = +28 | KCl | 2.2985 (2.2766, 2.3173) | 10.9938 (10.7795, 11.1438) |
| | NaCl | 2.276 (2.2664, 2.2860) | 10.6177 (10.5237, 10.7200) | | NaCl | 2.3053 (2.2909, 2.3209) | 10.8195 (10.6595, 10.9461) |
| | CaCl$_2$ | 1.7419 (1.7365, 1.7474) | 9.4547 (9.3311, 9.5447) | | CaCl$_2$ | 1.7587 (1.7549, 1.7622) | 9.7 (9.6082, 9.8055) |
| | MgCl$_2$ | 1.7535 (1.7505, 1.7568) | 9.6237 (9.5171, 9.7254) | | MgCl$_2$ | 1.7584 (1.7493, 1.7684) | 9.8941 (9.7292, 10.1252) |
| q = +56 | KCl | 2.4888 (2.4607, 2.5106) | 12.1131 (11.9218, 12.3005) | q = +84 | KCl | 2.5637 (2.5356, 2.5887) | 12.9305 (12.6435, 13.2986) |
| | NaCl | 2.4895 (2.4680, 2.5105) | 11.9343 (11.7345, 12.1722) | | NaCl | 2.5799 (2.5514, 2.6053) | 13.0737 (12.8194, 13.3270) |
| | CaCl$_2$ | 1.8741 (1.8700, 1.8781) | 10.6524 (10.5100, 10.7718) | | CaCl$_2$ | 1.9015 (1.8918, 1.9120) | 10.987 (10.9060, 11.0881) |
| | MgCl$_2$ | 1.8819 (1.8623, 1.8976) | 10.6579 (10.4874, 10.8363) | | MgCl$_2$ | 1.9327 (1.9152, 1.9504) | 11.2805 (11.1499, 11.4110) |

**Table S.17.** Same as Table S.16, but for the case of DND–NH$_2$.

| | Salt | $\tau^{(c)}$ / ps | $\tau^{(i)}$ / ps | | Salt | $\tau^{(c)}$ / ps | $\tau^{(i)}$ / ps |
|---|---|---|---|---|---|---|---|
| q = 0 | KCl | 2.286 (2.2834, 2.2886) | 10.4313 (10.3399, 10.5228) | q = +28 | KCl | 2.3238 (2.3046, 2.3456) | 11.0572 (10.9666, 11.2258) |
| | NaCl | 2.2925 (2.2771, 2.3047) | 10.7986 (10.6321, 11.0312) | | NaCl | 2.3119 (2.2992, 2.3214) | 10.9867 (10.8061, 11.1444) |
| | CaCl$_2$ | 1.7362 (1.7290, 1.7438) | 9.5468 (9.4784, 9.6224) | | CaCl$_2$ | 1.7707 (1.7632, 1.7766) | 9.8773 (9.8107, 9.9412) |
| | MgCl$_2$ | 1.7543 (1.7478, 1.7591) | 9.6061 (9.2989, 9.8174) | | MgCl$_2$ | 1.7821 (1.7723, 1.7927) | 10.003 (9.9203, 10.0621) |
| q = +56 | KCl | 2.3687 (2.3470, 2.3905) | 11.7944 (11.7216, 11.9098) | q = +84 | KCl | 2.5004 (2.4887, 2.5109) | 13.7065 (13.3743, 14.0386) |
| | NaCl | 2.3755 (2.3571, 2.3971) | 11.8577 (11.5739, 12.1612) | | NaCl | 2.485 (2.4518, 2.5183) | 14.72 (14.3417, 15.0983) |
| | CaCl$_2$ | 1.7953 (1.7881, 1.8043) | 10.343 (10.2374, 10.4452) | | CaCl$_2$ | 1.8355 (1.8283, 1.8435) | 11.434 (11.2581, 11.6049) |
| | MgCl$_2$ | 1.8127 (1.8068, 1.8186) | 10.4953 (10.3531, 10.6376) | | MgCl$_2$ | 1.8521 (1.8408, 1.8632) | 11.6365 (11.5435, 11.7261) |

**Table S.18.** Same as Table S.16, but for the case of DND–COOH.

| | Salt | $\tau^{(c)}$ / ps | $\tau^{(i)}$ / ps | | Salt | $\tau^{(c)}$ / ps | $\tau^{(i)}$ / ps |
|---|---|---|---|---|---|---|---|
| q = 0 | KCl | 2.2803 (2.2559, 2.3061) | 10.8194 (10.5260, 11.3353) | q = −28 | KCl | 2.2784 (2.2652, 2.2894) | 10.4739 (10.3223, 10.6408) |
| | NaCl | 2.2966 (2.2850, 2.3061) | 10.7663 (10.6478, 10.8915) | | NaCl | 2.2776 (2.2722, 2.2840) | 10.7999 (10.6108, 10.9891) |
| | CaCl₂ | 1.7377 (1.7304, 1.7469) | 9.5262 (9.4047, 9.6478) | | CaCl₂ | 1.7475 (1.7431, 1.7525) | 9.6252 (9.5160, 9.7121) |
| | MgCl₂ | 1.7465 (1.7387, 1.7535) | 9.6349 (9.4540, 9.7689) | | MgCl₂ | 1.7527 (1.7440, 1.7638) | 9.6198 (9.4847, 9.7333) |
| q = −56 | KCl | 2.2609 (2.2131, 2.2954) | 10.6699 (10.3367, 10.9008) | q = −84 | KCl | 2.2903 (2.2886, 2.2919) | 10.2784 (10.1875, 10.3694) |
| | NaCl | 2.2786 (2.2654, 2.2927) | 10.6328 (10.3468, 10.9226) | | NaCl | 2.2649 (2.2532, 2.2750) | 10.6159 (10.4875, 10.7086) |
| | CaCl₂ | 1.7458 (1.7388, 1.7515) | 9.504 (9.4366, 9.5749) | | CaCl₂ | 1.7375 (1.7276, 1.7465) | 9.5664 (9.5205, 9.6170) |
| | MgCl₂ | 1.7428 (1.7378, 1.7468) | 9.668 (9.5687, 9.7794) | | MgCl₂ | 1.7506 (1.7442, 1.7580) | 9.5821 (9.5246, 9.6386) |

**Table S.19.** Same as Table S.16, but for the case of DND–OH.

| | Salt | $\tau^{(c)}$ / ps | $\tau^{(i)}$ / ps |
|---|---|---|---|
| q = 0 | KCl | 2.2873 (2.2701, 2.3045) | 10.4815 (10.4699, 10.4931) |
| | NaCl | 2.2713 (2.2536, 2.2885) | 10.5693 (10.4172, 10.6794) |
| | CaCl₂ | 1.7424 (1.7290, 1.7554) | 9.555 (9.4417, 9.6683) |
| | MgCl₂ | 1.7496 (1.7406, 1.7623) | 9.7228 (9.6525, 9.7976) |

## References

1. Efron, B. & Tibshirani, R. An Introduction to the Bootstrap. (Chapman and Hall, 1993).